# Robust inference with GhostKnockoffs in genome-wide association studies

Xinran Qi[1], Michael E. Belloy[1], Jiaqi Gu[1], Xiaoxia Liu[1], Hua Tang[2]*, Zihuai He[1,3]*

* co-corresponding authors

[1]Department of Neurology and Neurological Sciences, Stanford University, Stanford, CA 94305, USA.

[2]Department of Genetics, Stanford University, Stanford, CA 94305, USA.

[3]Quantitative Sciences Unit, Department of Medicine, Stanford University, Stanford, CA 94305, USA.

# Correspondence to Xinran Qi (xinranqi@stanford.edu)

## Abstract

Genome-wide association studies (GWASs) have been extensively adopted to depict the underlying genetic architecture of complex diseases. Motivated by GWASs limitations' in identifying small effect loci to understand complex traits' polygenicity and fine-mapping putative causal variants from proxy ones, we propose a knockoff-based method which only requires summary statistics from GWASs and demonstrate its validity in the presence of relatedness. We show that GhostKnockoffs inference is robust to its input Z-scores as long as they are from valid marginal association tests and their correlations are consistent with the correlations among the corresponding genetic variants. The property generalizes GhostKnockoffs to other GWASs settings, such as the meta-analysis of multiple overlapping studies and studies based on association test statistics deviated from score tests. We demonstrate GhostKnockoffs' performance using empirical simulation and a meta-analysis of nine European ancestral genome-wide association studies and whole exome/genome sequencing studies. Both results demonstrate that GhostKnockoffs identify more putative causal variants with weak genotype-phenotype associations that are missed by conventional GWASs.

## Introduction

Genome-wide association studies (GWASs) have been extensively adopted to depict the underlying genetic architecture of complex phenotypes and diseases. As of now, the National Human Genome Research Institute-European Bioinformatics Institute (NHGRI-EBI) Catalog of human GWASs (Sollis et al., 2023) contains 6,566 publications, 302,340 single-nucleotide polymorphisms (SNPs), 552,116 top genotype-phenotype associations, and 65,590 open-source summary-level statistics. Despite the considerable success, conventional GWASs based on marginal association tests encountered two challenges in the post-GWAS era. First, while a large proportion of heritability remains unexplained, there is growing evidence that supports the high polygenicity of many complex traits. A multitude of small effect risk loci that currently lie below the stringent threshold that controls for family-wise error rate (FWER) can be informative to understand complex traits and enhance risk predictions. The more recent omnigenic model (Boyle et al., 2017) also supports lines of research from discovering "isolated" variant-gene-phenotype triplets to probing into pathways and regulations among variants with infinitesimal effect that influence complex phenotypes. Second, the marginal association tests identify abundant proxy variants located nearby due to their linkage disequilibrium (LD) with the causal ones. Researchers have developed second-stage conditional analysis and fine-mapping approaches to pinpoint credible sets of variants harboring the causal variant. However, these methods tend to focus on existing strong associations and lack systematic criteria for deducing the number of independent causal effects underlying association signals.

In recent years, model-X knockoffs inference (Candes et al., 2018) has been proposed as an alternative statistical framework to identify putative causal genetic variants for complex phenotypes. Complementary to GWASs testing for marginal genotype-phenotype association and controlling for FWER, knockoffs inference performs high-dimensional conditional independent test and controls for FDR, with corresponding null hypothesis "phenotype $Y_i$ is independent of genotype $G_{ij}$ given remaining genotypes $\boldsymbol{G}_{i,-j}$", where $i \in \{1, \cdots, n\}$ denotes sample index and $j \in \{1, \cdots, \mathrm{p}\}$. denotes variant index. The conditional independent test is done by generating knockoff counterparts mimicking the LD pattern of original genetic variants but are conditional independent of phenotypes given original ones. Knockoff counterparts act as negative controls in contrast to original genetic variants to account for the LD pattern. Previous studies show that the variants identified by knockoffs inference are more likely to be causal. The FDR control also allows for additional discoveries of variants with weaker effect that are missed by conventional GWASs, especially for complex traits with a polygenic/omnigenic genetic architecture.

While sample relatedness is a major confounder in recent large-scale genetic association studies, most knockoffs-based methods were proposed for studies with independent samples. Bates et al. (2020) (Bates et al., 2020) and Yang et al. (2022) (Yang et al., 2022) proposed methods for studies with known pedigree structure (e.g. studies with trio design). Sesia et al. (2021) (Sesia et al., 2021) proposed a method that controls for unknown diverse ancestries or familial relatedness by constructing knockoffs counterparts that retain the sample relatedness. However, the proposed knockoff construction is a highly nontrivial task and can be computationally intensive, which requires individual-level data to estimate haplotypes, estimate relatedness graph and partition related samples into independent identity-by-descent (IBD)-sharing families.

In this paper, we propose a simple and effective knockoffs-based method to account for sample relatedness, which only requires summary statistics from conventional GWASs. The proposed method is based on recently proposed GhostKnockoffs (He et al., 2022), which perform knockoffs inference without the need to generate individual-level knockoffs for hundreds of thousands of samples. We show in both theory and in empirical studies that GhostKnockoffs with input marginal Z-scores based on commonly used generalized linear mixed models can efficiently control for sample relatedness. In general, we show that GhostKnockoffs inference is robust to its input Z-scores as long as they are from valid marginal association tests and their correlations are consistent with the correlations among the corresponding genetic variants. This appealing property naturally extends GhostKnockoffs to many other useful settings in GWASs, such as the meta-analysis of multiple overlapping studies and studies based on test statistics deviated from conventional score tests.

We applied GhostKnockoff to the meta-analysis of nine European ancestral genome-wide association studies and whole exome/genome sequencing studies to select putative causal genetic variants for the Alzheimer's disease (AD). We showed that the proposed method identified 41 loci at FDR=0.1, including 23 loci that are missed by conventional marginal association tests with FWER control. We additionally leveraged external combined SNPs-to-gene strategy (cS2G) (Gazal et al., 2022) to further pinpoint functionally informed genes for the selected variants. Finally, we utilized single-cell RNA sequences (scRNAseq) consisting of 143,793 single-nucleus transcriptomes from two human brain components, hippocampus (9 AD cases and 8 controls) and cortex (4 AD cases and 4 controls) to validate genes mapped by the cS2G and other strategies.

## Results

### Method overview

We present a knockoff inference pipeline to perform genome-wide conditional independent tests in human genetic studies based on summary statistics (e.g., p-values). The proposed pipeline integrates recent advances in knockoff statistics, namely GhostKnockoffs, to improve power and to prioritize putative causal variants, generalized linear mixed effect model to control for sample relatedness, and meta-analysis strategies to aggregates multiple studies allowing for arbitrary sample overlap. The workflow is presented in **Figure 1**.

As shown in **Figure 1A**, our proposed method takes pre-calculated p-values as input. The p-value calculated can be from various types of marginal association tests whose embedded assumptions suit the data structure, such as generalized linear mixed effect model to account for sample relatedness, saddle point approximation for extreme case-control imbalance and meta-analysis that aggregates multiple studies. The p-values are then converted to corresponding Z-scores of a two-tailed significance test using the inverse standard Gaussian cumulative probability density function.

In contrast to conventional GWASs which perform marginal association tests to evaluate whether a variant is associated with the phenotype, our method utilizes a LD reference panel such as the Genome Aggregation Database (gnomAD) (Karczewski et al., 2020) to perform conditional tests evaluating whether a variant is associated with the phenotype conditioning on other variants in the same LD block (**Figure 2B**). The conditional test naturally reduces false positive findings due to LD confounding thereby prioritizing putative causal variants (He et al., 2022).

The pre-calculated Z-scores and LD reference panel serve as the input of the recently proposed GhostKnockoffs. As shown in **Figure 1C**, GhostKnockoffs generate multiple knockoff counterparts of the original Z-scores. The GhostKnockoffs' feature importance can be measured by the square of Z-scores and knockoff counterparts' Z-scores. The gap of feature importance scores between original genetic variants and corresponding knockoff counterparts quantifies association between variants and phenotypes and is later compared with the data-specific threshold controlling FDR to select putative casual variants.

We then leverage external functional information to further pinpoint mechanism-based target genes of the identified variants and to discover the underlying genetic structure of biological functions using the optimal combined SNP-to-gene (cS2G) strategy (**Figure 1D**) (Gazal et al., 2022). The cS2G strategy is a linear combination of constituent functionally informed S2G strategies (exon, promoter, GTEx fine-mapped *cis*-eQTL, eQTLGen fine-mapped blood *cis*-eQTL, EpiMap enhancer-gene linking, ABC, Cicero blood/basal; ordered from the highest to the lowest weight) to maximize heritability recall while guaranteeing heritability precision.

Finally, we perform downstream analysis of transcriptomics data to validate the enrichment of biological function of the mapped genes (**Figure 1E**) using single-cell RNA-sequencing data. In this particular data analysis for the Alzheimer's disease, we use 143,793 single-nucleus transcriptomes from 17 hippocampus (8 controls and 9 AD cases) and 8 cortex samples (4 controls and 4 AD cases) (Yang et al., 2021). We compare the enrichment analysis of single-cell transcriptomics data for AD-related genes identified by the cS2G and other strategies based on the proposed method and conventional GWASs variable selection results.

**Simulation results**

Our simulation studies aim to evaluate the robustness of GhostKnockoffs inference to various types of input Z-scores in terms of FDR and power, including: (1) Z-scores from generalized linear mixed models that account for sample relatedness; (2) Z-scores from a meta-analysis of multiple studies; (3) Z-scores from statistical tests that are deviated from usual score test. To mimic the real data scenario, we directly simulate genetic data based on the haplotype dataset from the R package *SKAT*, which mimics the LD structure of the European ancestry. For simulation studies that involve sample relatedness, we additionally simulate offspring genotypes using the gene dropping algorithm (MacCluer et al., 1986) to generate three-generation pedigrees. We then simulate quantitative and dichotomous phenotypes based on generalized linear models (generalized linear mixed model for simulation studies that involve sample relatedness). We present the simulation details in the "Methods" section.

For (1), we compare GhostKnockoffs with the second-order knockoffs that generates individual-level knockoffs per sample without the consideration of relatedness structure, referred to as "IndividualData knockoffs" in figures of the simulation results. The feature importance scores of two methods are set to be the same, square of Z-scores based on marginal association tests, to achieve unbiased comparison. We considered two types of marginal association tests: score test that ignores sample relatedness and score test from a generalized linear mixed model that correctly accounts for sample relatedness among the phenotypes. In **Supplementary Figure 6**, we present the QQ-plots of the two tests for the corresponding scenarios, which confirm the validity of the simulation setting. We present the simulation results in **Figure 2**.

We observe that for unrelated individuals, all methods control FDR and exhibit similar power. This confirms that they are theoretically equivalence in this scenario as shown by He et al. (He et al., 2022). For scenarios where the different individuals share correlated genotypes due to family relatedness, FDR inflation starts to appear. GhostKnockoffs inference with input Z-scores from generalized linear mixed model is the only method that has valid FDR control. The FDR inflation of the other three methods is more pronounced with higher level of sample relatedness and for scenarios in which genotypes and phenotypes are both related. The empirical results demonstrate that GhostKnockoffs, paired with commonly used generalized linear mixed model score test, can be naturally applied to genetic studies with sample relatedness to perform conditional independent tests with valid FDR control. We present additional comparisons at other FDR levels and other levels of sample relatedness in **Supplementary Figure 2-5**.

For (2), we compare GhostKnockoffs with input Z-scores from: a. meta-analysis of two independent studies (e.g. using the Fisher's combined probability test); b. mega-analysis by pooling samples from the two studies. We present the simulation results in **Figure 3A**. We observed that the two methods exhibit nearly equivalent FDR and power. The results demonstrate that GhostKnockoffs based on proper meta-analysis Z-scores guarantee rigorous FDR control and demonstrate similarly high power.

For (3), we compare GhostKnockoffs based on Z-scores from different types of association tests (score test, Wald test and likelihood ratio test). We present simulation results in **Figure 3B**. We observed that all three methods control FDR and exhibit similar power. GhostKnockoffs demonstrate robustness towards different association tests that yield Z-scores as the measurement of single-variant feature importance.

### Meta-analysis of Alzheimer's disease (AD) studies

We applied GhostKnockoffs to the meta-analysis Z-scores that aggregate nine European ancestral GWASs and whole exome/genome sequencing (WES/WGS) studies to select putative causal variants for AD. The calculation of the meta-analysis Z-scores accounting for sample overlapping can be found in the "Method" section. The nine studies include: (1) a genome-wide survival analysis of AD samples from the International Genomics of Alzheimer's Project (IGAP) (14,406 cases and 25,849 controls) (Huang et al., 2017); (2) a meta-analysis of GWASs in AD and AD-by-proxy based on parental diagnosis (71,880 cases and 383,378 controls) (Jansen et al., 2019); (3) a meta-analysis of GWASs in clinically diagnosed late-onset AD samples from the IGAP (21,982 cases and 41,944 controls) (Kunkle et al., 2019); (4) a meta-analysis of GWASs and GWASs-by-proxy in AD (53,042 cases and 355,900 controls) (Schwartzentruber et al., 2021) []; (5) GWASs of 32 AD cohorts (65,701 participants) (Le Guen et al., 2021); (6-7) WES analyses of the Alzheimer's Disease Sequencing Project (ADSP) by Bis et al. (5,740 cases and 5,096 controls) (Bis et al., 2020) and Le Guen et al. (6,008 cases and 5,119 controls) (Le Guen et al., 2021); (8-9) exome/genome-wide association analyses of ADSP (6,155 cases and 5,418 controls for ADSP WES, 3,584 cases and 2,949 controls for ADSP WGS) (Belloy et al., 2022). The meta-summary-level statistics consist of 9,195,254 common and low minor-allele-frequency (MAF) genetic variants with MAF>0.01.

The genetic variants are later matched with the LD reference panel from the gnomAD by chromosome, base pair position and reference/alternative alleles (Karczewski et al., 2020). We exclude variants if: (1) they do not show up in gnomAD; (2) they do not pass the gnomAD quality control; (3) they are multi-allelic; (4) they are in low complexity regions. To improve the power of conditional independent tests in presence of tightly correlated variants, we applied a single linkage hierarchical clustering with a correlation cutoff of 0.75 and use one representative variant per cluster for GhostKnockoffs inference. At FDR levels 0.05/0.1/0.2, we report the selected representative variants together with their neighboring variants in the same cluster which are (1) in high LD (0.75 or more correlated with the representative ones) and (2) of significant signal strength (absolute values of marginal Z-scores no less than those of corresponding selected representative ones) as the variable selection results. To pinpoint the mechanism-based gene that is most likely linked to an identified variant, we applied the cS2G and Open Targets V2G strategies. We also reported the proximal genes (nearest genes) for comparison.

We present GhostKnockoffs results in **Figure 4A** and conventional GWASs (at p-value threshold $5 \times 10^{-8}$) results in **Figure 4B**. The performance of GhostKnockoffs is compared with GWASs in terms of discoveries of independent loci, which are defined as loci that are at least 1Mb away from one another. The Manhattan plots present the identified independent loci, and each locus is annotated with the cS2G gene corresponding to the variant with the minimum p-value. In **Supplementary Figure 7A-B**, we present Manhattan plots based on a different cS2G mapping strategy, where each locus is annotated with the cS2G gene that appears most frequently.

We observed that GhostKnockoffs identified 25 loci at FDR=0.05, 41 loci at FDR=0.1 and 74 at FDR=0.2. GWASs identified 25 loci at p-value threshold $5 \times 10^{-8}$. At FDR level 0.1, GhostKnockoffs identified 1.64-fold more loci compared to GWASs. GhostKnockoffs achieve higher statistical power due to its error measurement as FDR, whereas GWASs rely on a more conservative error measurement as FWER. With more selected significant loci, GhostKnockoffs identify weak associations missed by GWASs that may explain additional AD phenotypic variance. We observed that the median number of clusters selected by GhostKnockoffs per locus was 1. Each cluster includes the variants in high LD with the representative variant and neighboring variants with relatively strong marginal associations. We noticed that there existed loci with variants in high LD (e.g. HLA) where GhostKnockoffs failed to distinguish the putative causal variants from the proxy variants. GhostKnockoffs tend to select a large cluster in such scenarios (**Figure 5**).

We considered the cS2G strategy as a gene mapping procedure to link genetic variants selected by GhostKnockoffs to putative functional genes. We observed that the proportion of selected variants that could be functionally mapped to a gene was 45.46% at FDR=0.05, 43.54% at FDR=0.1, and 42.69% at FDR=0.2, significantly higher than that of all background variants (23.57%). For GWASs, the percentage was 59.66% at p-value threshold $5 \times 10^{-8}$, which was higher than those of GhostKnockoffs due to its more stringent FWER control and exclusion of neighboring AD-proxy variants in high LD with the selected ones. **Supplementary Table 1** summarizes 41 independent loci associated with AD based on GhostKnockoffs at FDR=0.1, including variants' basic information, GhostKnockoffs statistics, and disease susceptibility genes identified by cS2G and V2G strategies, proximal genes and other enrichment analysis results.

### Enrichment analysis of single-cell transcriptomics data

We validated the genes implicated by GhostKnockoffs using differentially expressed gene (DEG) analysis of single-cell RNA sequences (scRNAseq) consisting of 143,793 single-nucleus transcriptomes from two human brain components, hippocampus (9 AD cases 8 controls) and cortex (4 AD cases and 4 controls). We performed the DEG analysis using R package *Seurat*, stratified by 14 cell types (veinous endothelial cell, T cell, smooth muscle cell, pericyte, capillary endothelial cell, arterial endothelial cell, oligodendrocyte, perivascular fibroblast, ependymal cell, microglia, astrocyte, oligodendrocyte progenitor cell, meningeal fibroblast, neuron), adjusting for age, batch effect, cellular detection rate and relatedness within samples.

We present the scRNAseq DEG analysis results of cS2G genes based on GhostKnockoffs at FDR levels 0.05/0.1/0.2 and conventional GWASs in **Supplementary Figure 8A-D**. The volcano plots demonstrate genes' differential expression among 14 cell types in terms of (1) $\log_2$(fold change) of average expression between AD cases and controls, (2) $-\log_{10}$(p-value) of DE testing. Two thresholds, 0.05 and Bonferroni adjustment 0.05/(number of genes with expression measurements), are used to determine DEGs. **Figure 6A-C** compare proportions of differentially expressed cS2G, V2G and proximal genes based on GhostKnockoffs at FDR levels 0.05/0.1/0.2 and GWASs. We could see that GhostKnockoffs and GWASs enrich AD-related genes since their proportions of DEGs are constantly higher than the baseline, which is defined as the proportion of DEGs among all 23,537 background genes from the scRNAseq data, across all cell types. **Figure 6D** shows that proportions of differentially expressed cS2G, V2G and proximal genes in at least one cell type based on GhostKnockoffs and conventional GWASs are generally higher than the baseline. At FDR level 0.1, 55/79=69.62% cS2G genes (or 53/73=72.6% proximal genes, 58/75=77.33% V2G genes) based on GhostKnockoffs variable selection results are differentially expressed in at least one cell type and are significantly higher than the baseline 12,514/23,537=53.17%. **Supplementary Table 2-4** show detailed counts and percentages of differentially expressed genes identified by cS2G/V2G strategy and proximal genes based on GhostKnockoffs and conventional GWASs variable selection results. We see that GhostKnockoffs capture more weak associations of AD-related variants, whose disease susceptibility genes identified by the cS2G/V2G strategy or proximal genes are of higher magnitude compared to conventional GWASs. Interestingly, both methods demonstrate similar proportions of DEG enrichment in scRNAseq analysis.

## Discussions

We have proposed a simple and effective knockoffs-based method to account for sample relatedness, which only requires summary statistics from conventional GWASs. We show in both theory and in empirical studies that GhostKnockoffs inference is robust to cryptic relationship or overlap in the underlying study populations, as long as the input summary statistics are from valid marginal association tests and their correlations are consistent with the correlations among the corresponding genetic variants. These results generalize the application of GhostKnockoffs to broader GWASs settings, such as the meta-analysis of multiple overlapping studies and studies based on association test statistics deviated from score tests.

We applied GhostKnockoffs to the meta-analysis of nine European ancestral GWASs and WES/WGS studies to select putative causal genetic variants for the Alzheimer's disease. GhostKnockoffs identified much more loci that were missed by conventional GWASs. We also leveraged external SNPs-to-gene strategies (cS2G, Open Targets V2G and proximal genes) to further pinpoint functionally informed genes for the selected variants. Finally, we performed downstream analysis of transcriptomics data to validate the enrichment of biological function of the mapped genes. Both simulation results and the meta-analysis of AD studies demonstrate that GhostKnockoffs improve the power to identify more putative causal variants with weak genotype-phenotype associations and more mechanism-based genes.

One limitation of GhostKnockoffs is that there are genetic regions with variants in high LD and GhostKnockoffs fail to distinguish the putative causal variants from proxy ones. Since variants high LD decrease the power to identify causal ones, variants are then grouped into hierarchical clusters first and GhostKnockoffs are applied to select the representative variant from each cluster. In this case, even though theoretically, GhostKnockoffs adjust for the LD by preserving the correlation structure of genotypes among knockoff counterparts, it fails to identify putative causal variants within the same hierarchical cluster. To avoid missing the "real" causal variants, the selected representative variants and neighboring ones from the same cluster are treated as GhostKnockoffs variable selection results, meaning that many proxy variants still waits for further step of fine-mapping. We are interested in exploring alternative knockoffs inference methods to integrate biological annotations and other supportive information as an additional layer of counterparts for the original variants besides knockoffs to further finemap putative causal variants from proxy ones.

## Figures

**Figure 1: Method overview.** Schematic of the proposed GhostKnockoffs approach based on summary statistics. **A.** Input Z-scores can be derived from GWAS summary statistics **B.** Linkage disequilibrium (LD) matrix can be obtained from a reference panel. **C.** GhostKnockoffs inference using Z-scores and LD reference panel as input. **D.** Pinpoint mechanism-based target genes of the identified variants. **E.** Downstream analysis of transcriptomics data to validate the enrichment of biological function of the mapped genes.

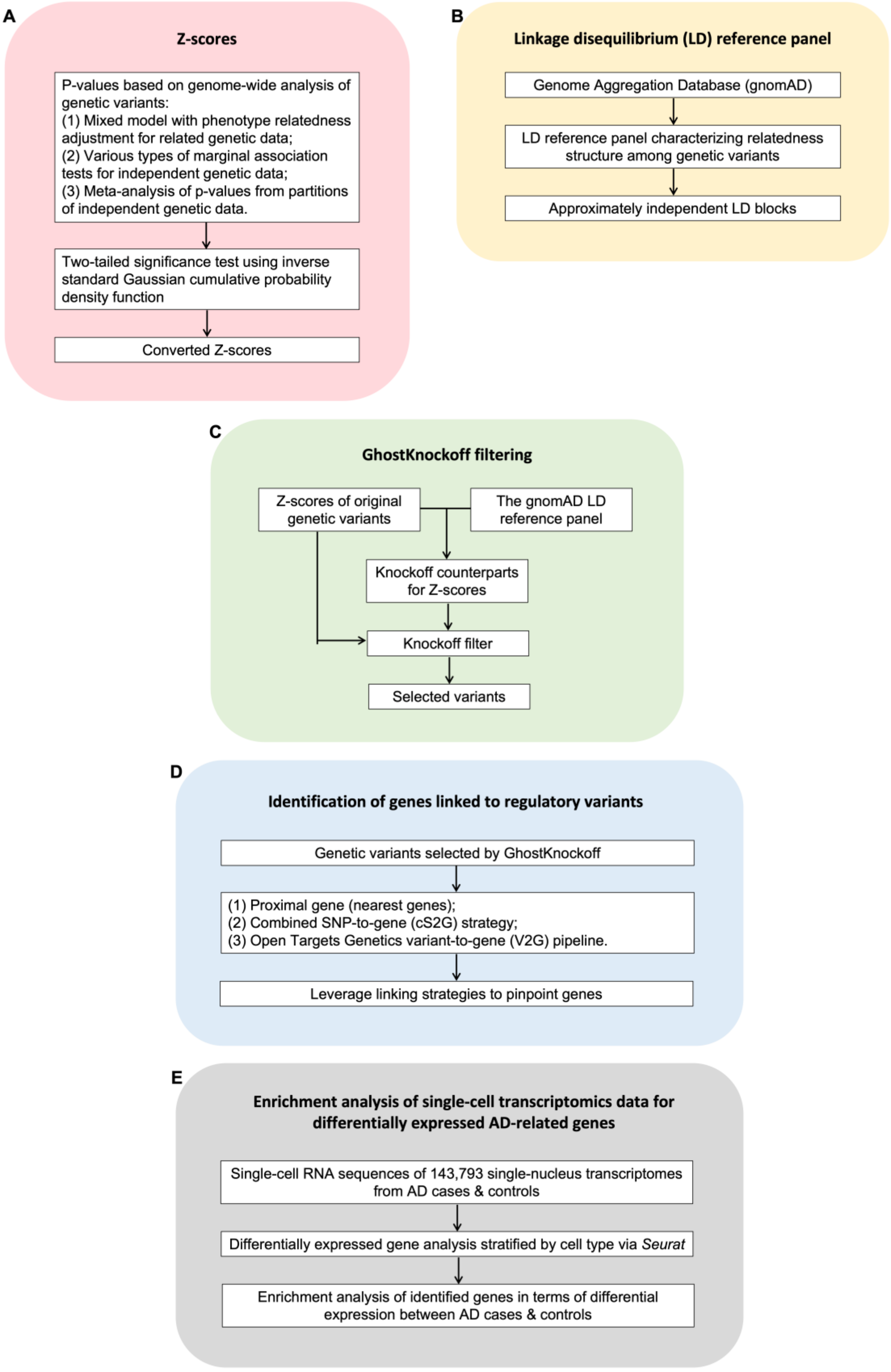

A
Z-scores
P-values based on genome-wide analysis of genetic variants:
(1) Mixed model with phenotype relatedness adjustment for related genetic data;
(2) Various types of marginal association tests for independent genetic data;
(3) Meta-analysis of p-values from partitions of independent genetic data.
Two-tailed significance test using inverse standard Gaussian cumulative probability density function
Converted Z-scores
B
Linkage disequilibrium (LD) reference panel
Genome Aggregation Database (gnomAD)
LD reference panel characterizing relatedness structure among genetic variants
Approximately independent LD blocks
C
GhostKnockoff filtering
Z-scores of original genetic variants
The gnomAD LD reference panel
Knockoff counterparts for Z-scores
Knockoff filter
Selected variants
D
Identification of genes linked to regulatory variants
Genetic variants selected by GhostKnockoff
(1) Proximal gene (nearest genes);
(2) Combined SNP-to-gene (cS2G) strategy;
(3) Open Targets Genetics variant-to-gene (V2G) pipeline.
Leverage linking strategies to pinpoint genes
E
Enrichment analysis of single-cell transcriptomics data for differentially expressed AD-related genes
Single-cell RNA sequences of 143,793 single-nucleus transcriptomes from AD cases & controls
Differentially expressed gene analysis stratified by cell type via Seurat
Enrichment analysis of identified genes in terms of differential expression between AD cases & controls

**Figure 2: Empirical simulation studies of GhostKnockoffs for different levels of sample relatedness at FDR level 0.1.** We present FDR and power estimation of GhostKnockoffs and the second-order knockoffs based on 1000 replicates of 10,000 samples for quantitative and dichotomous phenotypes. GhostKnockoff, mixed model score test/score test: multiple GhostKnockoffs based on Z-scores converted from p-values of mixed effect model score test by adjusting kinship among phenotypes/score test without phenotype kinship adjustment. IndividualData knockoff, mixed model score test/score test: multiple second-order knockoffs based on Z-scores converted from p-values of mixed effect model score test by adjusting kinship among phenotypes/score test without phenotype kinship adjustment.

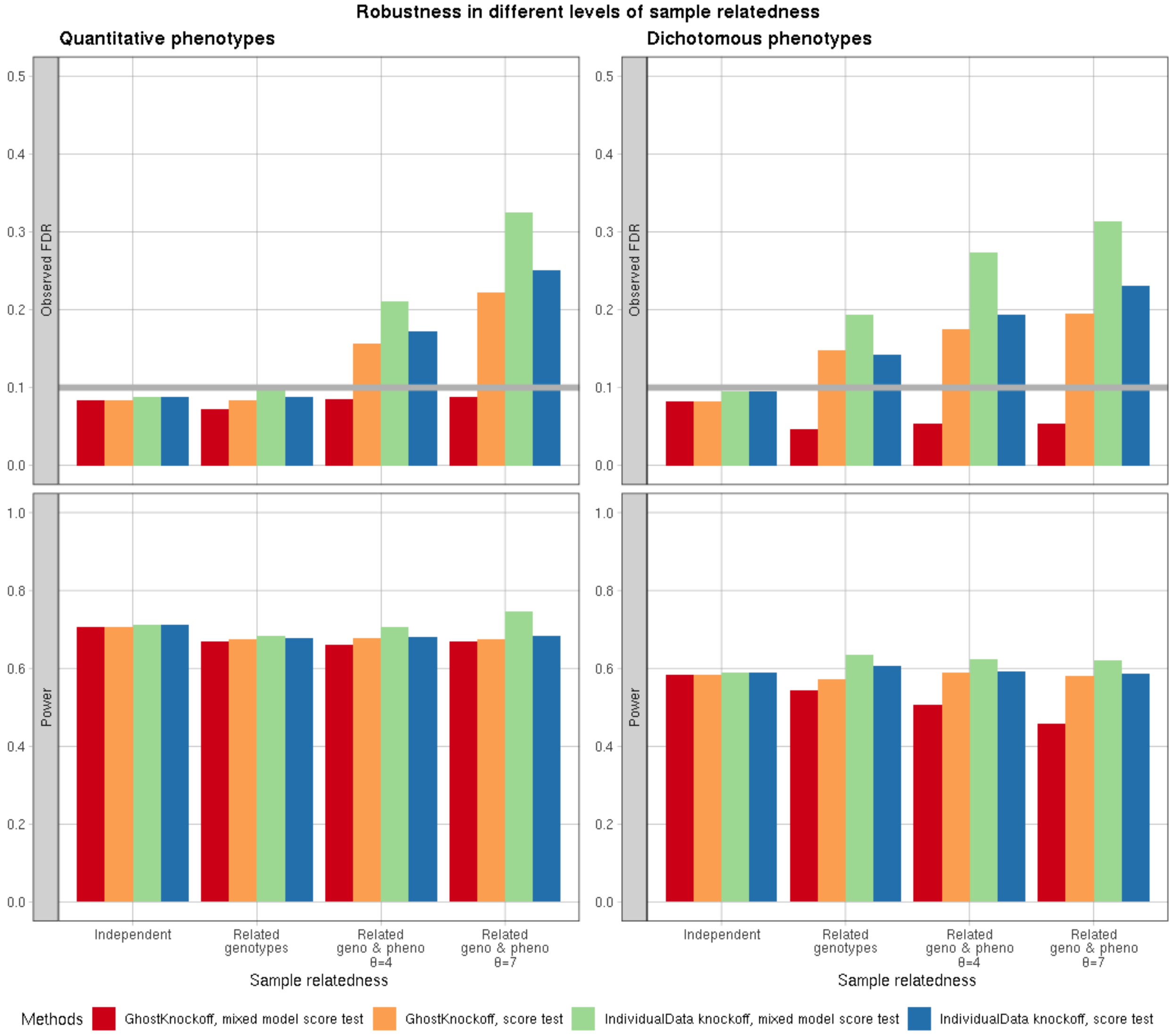

**Figure 3: Empirical simulation studies of GhostKnockoffs for meta-analysis of multiple studies and different types of association tests. A.** FDR and power estimation of GhostKnockoffs with input Z-scores from: a. meta-analysis of two independent studies; b. mega-analysis by pooling samples from the two studies. Comparisons are based on 1000 replicates of 10,000 independent samples for quantitative and dichotomous phenotypes. GhostKnockoff, score test/meta score test (Fisher's method): multiple GhostKnockoffs based on p-values of score test applied to samples pooled from the two studies and meta-analysis of score test p-values calculated separately from two independent studies using Fisher's method. **B.** FDR and power estimation of GhostKnockoffs with input Z-scores from different types of association

tests. Comparisons are based on 1000 replicates of 10,000 independent samples for quantitative and dichotomous phenotypes. GhostKnockoff, likelihood ratio test/score test/Wald test: multiple GhostKnockoffs with input Z-scores converted from p-values of likelihood ratio test/score test/Wald test applied to independent samples.

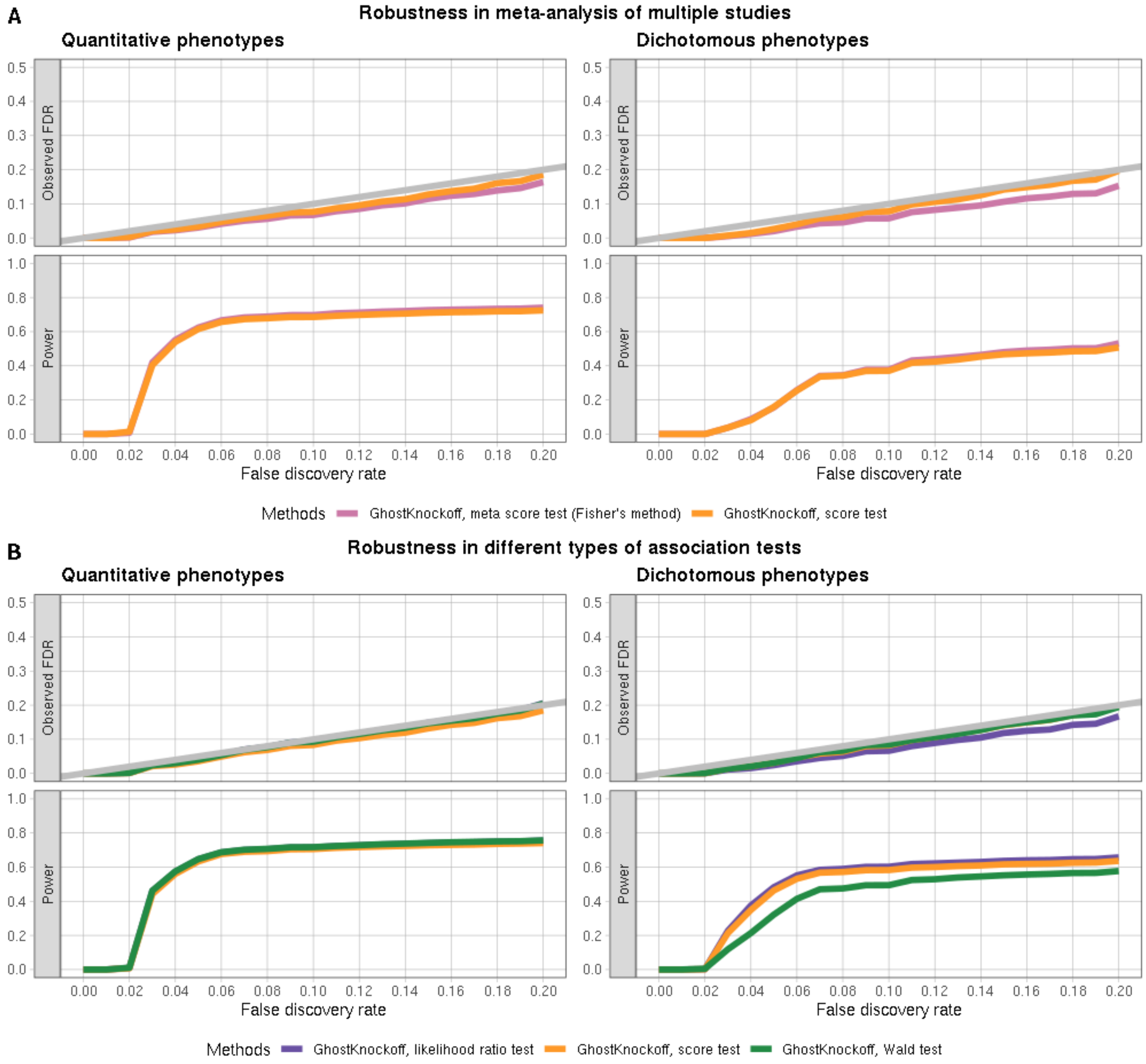

**Figure 4: Meta-analysis of Alzheimer's disease studies.** We present Manhattan plots of GhostKnockoffs and conventional GWASs applied to the meta-analysis Z-scores that aggregate nine European ancestral GWASs and WES/WGS studies. Each locus is annotated with the cS2G gene corresponding to the variant with the minimum p-value. The variant density of each independent locus (number of variants per 1Mb) is shown at the bottom of plots. **A.** Manhattan plot of $W$ statistics (truncated at 100) based on GhostKnockoffs at FDR levels 0.05 (red horizontal dashed line) and 0.1 (black horizontal dashed line). **B.** Manhattan plot of $-\log_{10}$(p-value) (truncated at 50) based on conventional GWASs at p-value threshold $5 \times 10^{-8}$ (black horizontal dashed line).

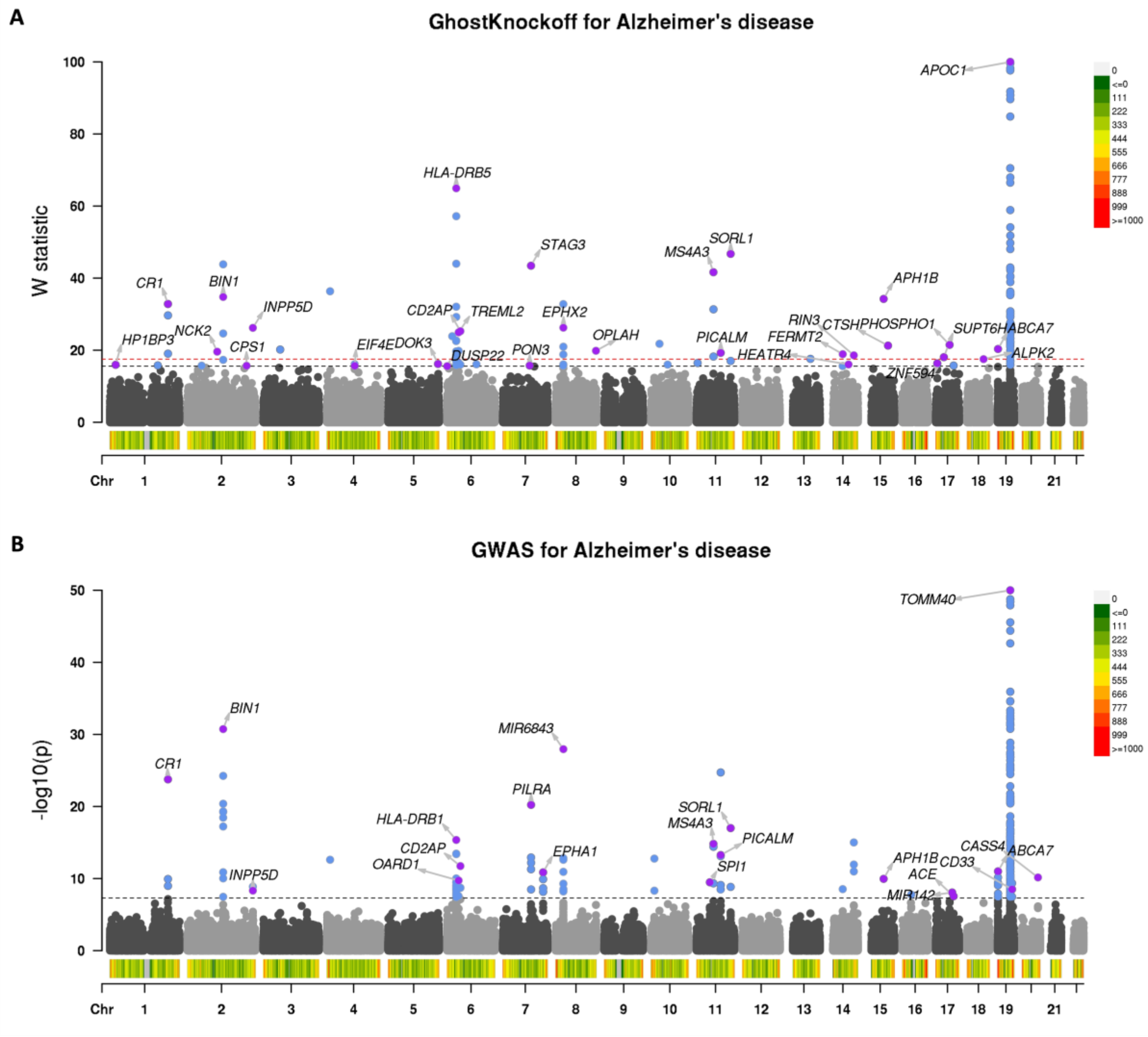

**Figure 5: Genetic variants of hierarchical clusters selected by GhostKnockoffs. A.** We present scatter plots of clusters selected by GhostKnockoffs at FDR levels 0.05/0.1/0.2. The x-axis denotes the number of variants of each selected cluster. The y-axis denotes the signal strength ($\tau$ denotes the difference between the highest importance score, measured by the square of Z-scores, and the median of remaining importance scores) of the selected representative variant of each cluster. The signal strength $\tau$ is truncated at 100 for better visualization. Red points denote clusters consisting of no less than 35 variants with corresponding gene region annotated above. **B.** We present histograms of the number of variants per cluster selected by GhostKnockoffs at FDR levels 0.05/0.1/0.2. The blue vertical dashed line denotes the average number of variants per cluster.

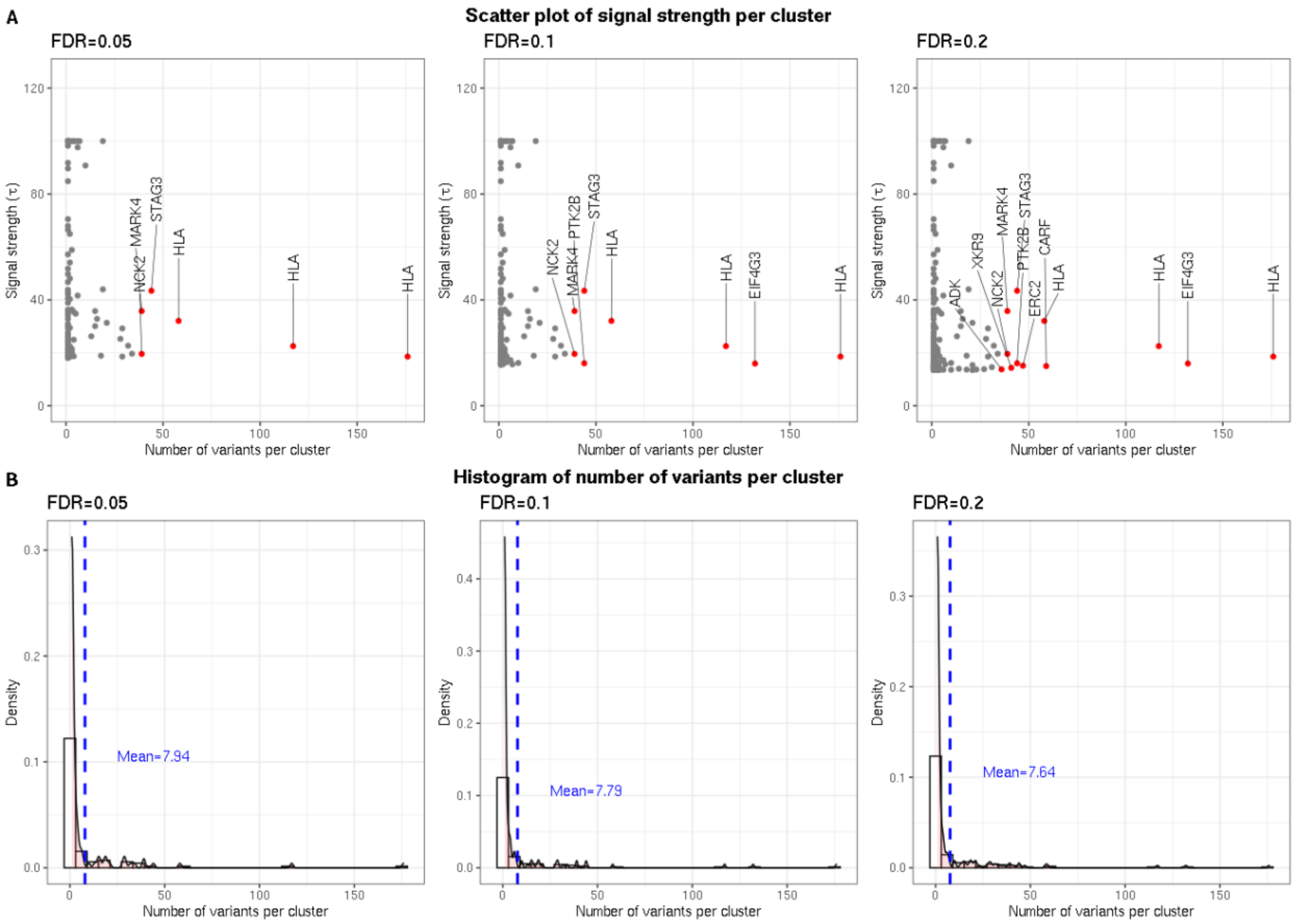

**Figure 6: Enrichment analysis of single-cell transcriptomics data for identified AD-related genes.** We present differentially expressed gene (DEG) analysis of single-cell RNA sequences consisting of 143,793 single-nucleus transcriptomes for AD-related genes identified by cS2G/V2G strategy and proximal genes based on GhostKnockoffs and conventional GWASs variable selection results to validate their signal enrichment. **A.** We present proportions of DEGs identified by the cS2G strategy stratified by 14 cell types. Genes are classified as DEGs if their DE analysis raw p-values are smaller than 0.05. **B.** We present proportions of differentially expressed proximal genes stratified by 14 cell types. **C.** We present proportions of DEGs identified by the V2G strategy stratified by 14 cell types. **D.** We present proportions of DEGs identified by three based on GhostKnockoffs (FDR=0.05/0.1/0.2), GWASs and all background genes. Genes are classified as DEG if any of the 14 cell types' DE analysis raw p-values is smaller than the 0.05.

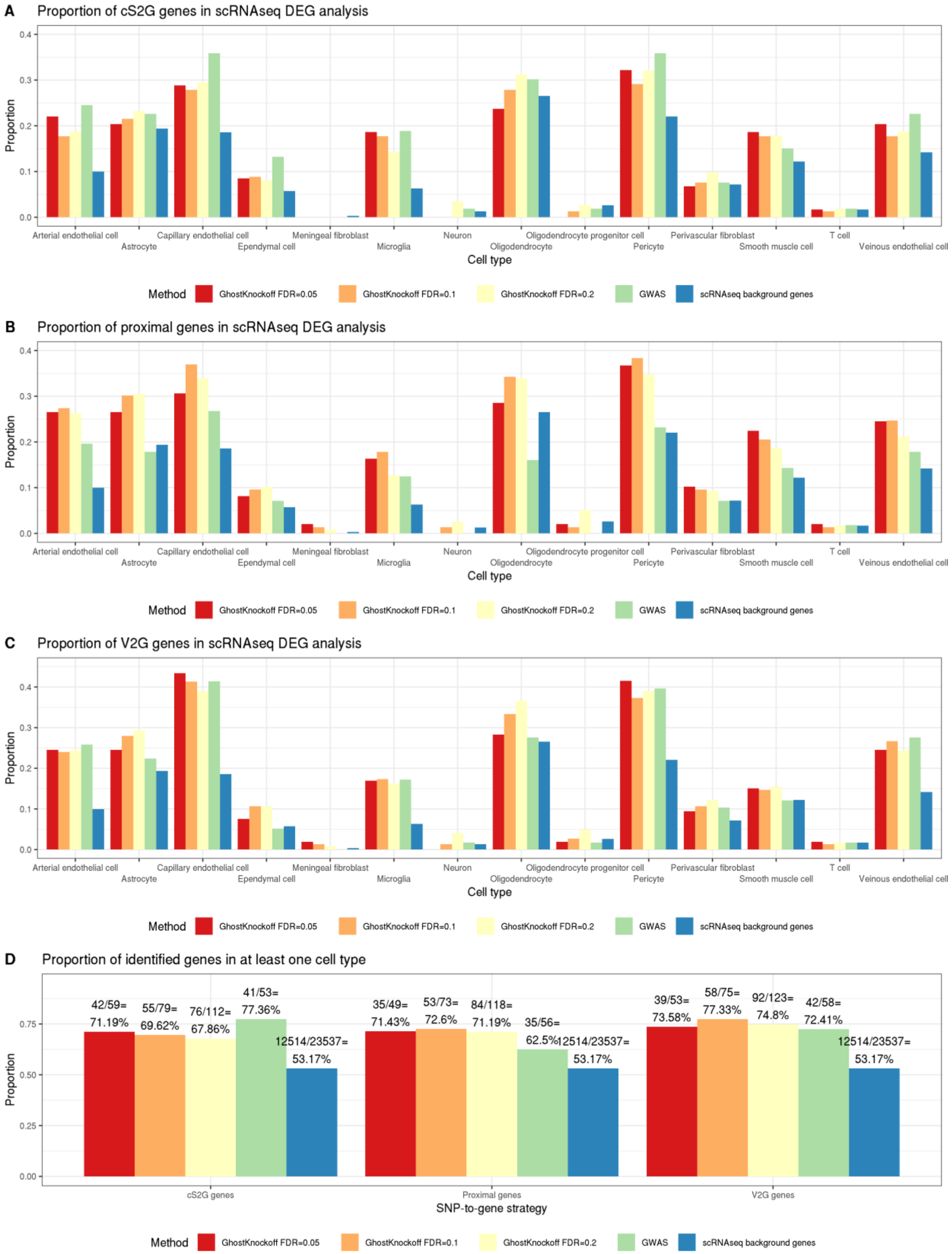
A
Proportion of cS2G genes in scRNAseq DEG analysis
B
Proportion of proximal genes in scRNAseq DEG analysis
C
Proportion of V2G genes in scRNAseq DEG analysis
Proportion
Cell type
Arterial endothelial cell
Astrocyte
Capillary endothelial cell
Ependymal cell
Meningeal fibroblast
Microglia
Neuron
Oligodendrocyte
Oligodendrocyte progenitor cell
Pericyte
Perivascular fibroblast
Smooth muscle cell
T cell
Veinous endothelial cell
Method
GhostKnockoff FDR=0.05
GhostKnockoff FDR=0.1
GhostKnockoff FDR=0.2
GWAS
scRNAseq background genes
D
Proportion of identified genes in at least one cell type
42/59= 71.19%
55/79= 69.62%
76/112= 67.86%
41/53= 77.36%
12514/23537= 53.17%
35/49= 71.43%
53/73= 72.6%
84/118= 71.19%
35/56= 62.5%
12514/23537= 53.17%
39/53= 73.58%
58/75= 77.33%
92/123= 74.8%
42/58= 72.41%
12514/23537= 53.17%
cS2G genes
Proximal genes
V2G genes
SNP-to-gene strategy

## Methods

### Method overview

### [Corresponding to Figure 1A] P-value calculation via generalized linear mixed model (GLMM) and score test

While the proposed method can take p-values from various types of marginal association tests as input, we use the commonly used generalized linear mixed model as an example to describe the method. Consider a study with $n$ possibly related participants, $p$ genetic variants and $q$ other covariates. The relationship between participants is not assumed to be known. For the $i^{th}$ individual, the scalar $Y_i$ denotes its phenotype measurement (either quantitative or dichotomous); the $1 \times p$ row vector $\boldsymbol{G_i} = (G_{i1}, \cdots, G_{ip}) \in \{0,1,2\}^p$ denotes genotypes of genetic variants; the $1 \times q$ row vector $\boldsymbol{X_i} = (X_{i1}, \cdots, X_{iq})$ denotes other covariates. The GLMM for quantitative/dichotomous phenotype is:

$$g(\mu_i) = \boldsymbol{X_i \alpha} + \boldsymbol{G_i \beta} + b_i,$$

where $\mu_i = \mathrm{E}(y_i | \boldsymbol{X_i}, \boldsymbol{G_i}, b_i)$ denotes the expectation of the $i^{th}$ phenotype $y_i$, conditional on covariate $\boldsymbol{X_i}$, genotype $\boldsymbol{G_i}$ and random effect $b_i$. The $p \times 1$ column vector $\boldsymbol{\beta_i} = (\beta_{i1}, \cdots, \beta_{ip})^T$ denotes genotype effects. The $q \times 1$ column vector $\boldsymbol{\alpha_i} = (\alpha_{i1}, \cdots, \alpha_{ip})^T$ denotes fixed covariate effects plus intercept. The link function is set to be $g(\mu) = \mu$ for a quantitative normally distributed phenotype and $g(\mu) = \mathrm{logit}(\mu)$ for a dichotomous phenotype. The $n \times 1$ column vector $\boldsymbol{b} = (b_i)^T \sim N(\boldsymbol{0}, \theta\boldsymbol{\Phi})$ denotes random effect with the variance component parameter $\theta$ and the $n \times n$ matrix $\boldsymbol{\Phi}$ measuring sample relatedness. The scalar $\epsilon_i \sim N(0, \mathrm{Var}(\epsilon))$ denotes the residual of the linear model.

The Z-scores based on single-variant score test p-values, denoted as $Z_{score,k} = sign(\beta_k)\Phi^{-1}\left(\frac{p-value_k}{2}\right)$, (see Supplementary Materials for derivations) are utilized for the next steps.

### [Corresponding to Figure 1B] The gnomAD LD reference panel to facilitate conditional test

Our method utilizes the European LD reference panel from the Genome Aggregation Database (gnomAD) (Karczewski et al., 2020) to perform conditional tests. The gnomAD estimates LD more accurately by aggregating worldwide large-scale sequencing projects' exome and genome sequence data to yield summary data, which provides substantial power guarantee for human genetic studies. It surpasses the 1000 Genomes Project whose majority samples of exome sequence data are included in gnomAD and hence makes more robust estimates of the LD reference panel. As shown in **Figure 1B**, the gnomAD provides summary data from "unrelated" individuals, whose first and second-degree relatives are excluded to avoid rare variants' allele frequency inflation, as the reference for coding region analysis with substantial power increase. With sample and variant quality control, the LD reference panel characterizes the relatedness structure among genetic variants more robustly compared to most genome database.

### [Corresponding to Figure 1C] GhostKnockoffs inference based on Z-scores

As opposed to typical model-X knockoffs which require individual-level genotype data, He et al. (2022) proposed the GhostKnockoffs which directly use the Z-scores as input to generate multiple knockoff counterparts by

$$\boldsymbol{\widetilde{Z}_{score}} = \boldsymbol{PZ_{score}} + \boldsymbol{E}, \boldsymbol{E} \sim \boldsymbol{N}(\boldsymbol{0}, \boldsymbol{V}) \quad (1)$$

where the $pM \times 1$ column vector $\boldsymbol{\widetilde{Z}_{score}} = (\boldsymbol{\widetilde{Z}^{m}_{score}})_{\boldsymbol{1 \le m \le M}}$ consists of Z-scores for $M$ knockoff counterparts per genetic variant. The $pM \times p$ matrix $\boldsymbol{P}$ and $pM \times pM$ matrix $\boldsymbol{V}$ are defined by the input LD matrix such that the inference based on summary statistics is equivalent to the multiple second-order knockoffs inference based on individual-level data for independent samples (He et al., 2022). Intuitively, matrix $\boldsymbol{P}$ serves as a projection matrix based on LD such that $\boldsymbol{\widetilde{Z}_{score}}$ quantifies the indirect effect through correlation with other variants. The corresponding feature importance scores $\boldsymbol{T}^m = \left(\widetilde{\boldsymbol{Z}}_{\boldsymbol{score}}^m\right)^2$ are compared with $\boldsymbol{T} = (\boldsymbol{Z}_{\boldsymbol{scores}})^2$, refered to as

$$\boldsymbol{W} = \left(\boldsymbol{T} - \underset{1 \le m \le M}{\text{median}} \boldsymbol{T}^{\boldsymbol{m}}\right) I_{\boldsymbol{T} \ge \underset{1 \le m \le M}{\max} \boldsymbol{T}^{\boldsymbol{m}}},$$

to quantify the association between variants and the phenotype, conditioning on other variants in the same LD block. Based on the user-specified FDR level, a threshold $\tau$ is chosen and genetic variants whose $W > \tau$ are selected by GhostKnockffs.

While the GhostKnockoffs method is originally proposed for independent samples, we demonstrate its robustness to input summary statistics via extensive simulation studies. Specifically, we show that the GhostKnockoffs inference remains valid if the input p-values are correctly calculated.

### Simulate unrelated & related genotypes

We simulate unrelated and related genotypes based on the haplotype dataset from the R package *SKAT*, which consists of 10,000 haplotypes over 200K BP region and is generated from the calibration coalescent model (COSI) software and mimics the LD structure of the European ancestry (Schaffner et al., 2005). Unrelated genotypes are simulated based on randomly sampled haplotypes. Related genotypes are simulated using the three-generation pedigree, as shown in **Supplementary Figure 1**. For each family that consists of 10 family members, two unrelated grandparents are simulated based on randomly sampled haplotypes. Their offspring genotypes (the second-generation dependent parents) are simulated using the gene dropping algorithm. Given that the *SKAT* haplotypes dataset consists of haplotypes over 200K base pair region, the recombination rate is set to be 0 while simulating offspring genotypes. Two other unrelated second-generation parents are randomly sampled from the *SKAT* haplotypes dataset similarly to pair with the second-generation dependent parents correspondingly. The third-generation offspring genotypes are simulated in a similar way using the gene dropping algorithm.

We simulate genotypes of (a) 10,000 unrelated individuals and b. 1000 families according to the pedigree in **Supplementary Figure 1.** 1000 genetic variants are randomly selected, out of which variants whose MAF greater than 0.01 are kept for the robust calculation of marginal association test p-values. We apply the single linkage hierarchical clustering to variants and group them to clusters so that there are no two clusters having cross-correlation above 0.75. A representative variant with the biggest sum of absolute values of within-correlations is selected per cluster for simulation study.

### Simulate independent & related phenotypes

We randomly select 10 causal variants to simulate quantitative and dichotomous phenotypes. The absolute value of regression coefficient for the $j^{th}$ causal variant is $|\beta_j| = \sqrt{\frac{a}{10 \times \text{Var}(\boldsymbol{G}_{\cdot j})}}$, where $j \in \{1, \cdots, 1000\}$. The hyperparameter $a$ denotes the variance explained by causal variants and is a constant. For quantitative phenotypes, $a = 1$ and for dichotomous traits, $a = 2.5$. We define that half of causal variants show protective effect (negative sign of $\beta_j$) and the other half show risk effect (positive sign $\beta_j$).

***Independent phenotypes*** We simulate independent phenotypes based on unrelated genotypes $\boldsymbol{G}$ as follows.

Linear fixed effect model for quantitative phenotype $Y_i$:

$$Y_i = X_{i1} + \boldsymbol{G_i \beta} + \epsilon_i^Q.$$

Logistic fixed effect model for dichotomous phenotype $Y_i$:

$$\text{logit}(\pi_i) = \beta_0 + X_{i1} + \boldsymbol{G_i \beta} + \epsilon_i^D,$$

where $\boldsymbol{G_i} = (G_{ij})^T$ denotes genotypes for the $i^{th}$ individual, $i \in \{1, \cdots, 10000\}$. $X_{i1} \sim N(0,1)$ denotes the fixed effect covariate. $\epsilon_i^Q \sim N(0, \sqrt{8})$ denotes the residual of the linear model. $\epsilon_i^D \sim N(0,1)$ measures the variation due to unobserved covariates of the logistic model. $X_{i1}$, $\epsilon_i^Q$ and $\epsilon_i^D$ are mutually independent. $\beta_0$ denotes the intercept of the logistic model corresponding to prevalence being 10%. The link function is $\text{logit}(\pi_i) = \log\left(\frac{\pi_i}{1-\pi_i}\right)$. The $\pi_i = \Pr(Y_i = 1|X_{i1}, \boldsymbol{G_i})$ denotes the conditional mean of $Y_i$ given covariate $X_{i1}$ and genotypes $\boldsymbol{G_i}$.

***Related phenotypes*** We simulate related phenotypes based on related genotypes as follows.

Linear mixed effect model for quantitative phenotype $Y_i$:

$$Y_i = X_{i1} + \boldsymbol{G_i}\boldsymbol{\beta} + b_i + \epsilon_i^Q.$$

Logistic mixed effect model for dichotomous phenotype $Y_i$:

$$\text{logit}(\pi_i) = \beta_0 + X_{i1} + \boldsymbol{G_i}\boldsymbol{\beta} + b_i,$$

where $b_i$ denotes the random effect and $\boldsymbol{b} \sim N(0, \boldsymbol{\Sigma})$. The $\boldsymbol{\Sigma} = \theta\boldsymbol{\Phi}$ denotes the variance covariance matrix of random effects, where $\theta$ is the variance component parameter and $\boldsymbol{\Phi}$ denotes the kinship matrix corresponding to the three-generation-pedigree.

$$\boldsymbol{\Phi} = \begin{bmatrix} 1 & 0 & 0.5 & 0.5 & 0 & 0 & 0.25 & 0.25 & 0.25 & 0.25 \\ 0 & 1 & 0.5 & 0.5 & 0 & 0 & 0.25 & 0.25 & 0.25 & 0.25 \\ 0.5 & 0.5 & 1 & 0.5 & 0 & 0 & 0.5 & 0.5 & 0.25 & 0.25 \\ 0.5 & 0.5 & 0.5 & 1 & 0 & 0 & 0.25 & 0.25 & 0.5 & 0.5 \\ 0 & 0 & 0 & 0 & 1 & 0 & 0.5 & 0.5 & 0 & 0 \\ 0 & 0 & 0 & 0 & 0 & 1 & 0 & 0 & 0.5 & 0.5 \\ 0.25 & 0.25 & 0.5 & 0.25 & 0.5 & 0 & 1 & 0.5 & 0.125 & 0.125 \\ 0.25 & 0.25 & 0.5 & 0.25 & 0.5 & 0 & 0.5 & 1 & 0.125 & 0.125 \\ 0.25 & 0.25 & 0.25 & 0.5 & 0 & 0.5 & 0.125 & 0.125 & 1 & 0.5 \\ 0.25 & 0.25 & 0.25 & 0.5 & 0 & 0.5 & 0.125 & 0.125 & 0.5 & 1 \end{bmatrix}.$$

For linear mixed model, we set the sum of variance due to residual $\epsilon^Q$ and random effects $\boldsymbol{b}$ as a constant value. Therefore, $\epsilon_i^Q \sim N\left(0, \sqrt{\text{Var}(\epsilon^Q)}\right)$, where $\text{Var}(\epsilon^Q) = 8 - \text{Var}(\boldsymbol{b})$ so that we guarantee the power at a similar level while changing the value of variance component parameter $\theta \in \{1,4,7\}$. For logistic mixed effect model, the variance component parameter $\theta$ also varies within the range $\theta \in \{1,4,7\}$.

**Case control family study for related dichotomous phenotypes**

We design three case control family studies for dichotomous phenotypes to compare GhostKnockoffs with the second-order knockoffs under different levels of sample relatedness. The parameter $K = \left(\frac{\sum_{i,j} d(i,j) \times f(i,j)}{\#\text{ of } (i,j) \text{ pairs in family}}\right) / (\#\text{ of families})$ measures the sample relatedness of phenotypes from each sampling scheme. For each pair of individuals, $d(i,j) = 1$ if both are cases/controls and $d(i,j) = 0$ otherwise; $f(i,j) = 1$ if both are from the same family and $f(i,j) = 0$ otherwise.

The following three sampling schemes are based on related genotypes simulated using the gene dropping algorithm and related phenotypes simulated using the logistic mixed effect model, which we abbreviate as the simulated foundation data. Each sample scheme has 10,000 individuals.

***Scheme A*** We randomly select 5000 cases ($Y = 1$) and 5000 controls ($Y = 0$) from the simulated foundation data. ($K$=0.383)

***Scheme B*** We randomly select 500 case families (number of cases greater than or equal to 1 per family) and 500 control families (0 cases per family) from the simulated foundation data. ($K$=0.808)

***Scheme C*** We select 5000 cases by including all cases from case families and randomly select 5000 controls from all control families based on the simulated foundation data. ($K$=1)

**Empirical FDR and power**

We applied GhostKnockoffs and the second-order knockoffs to the simulation data. We ran 1000 replicates and estimated empirical FDR and power. The empirical FDR is estimated by the proportion of false positive variants among all selected ones. The empirical power is estimated by the proportion of selected variants among all causal ones.

**Meta-analysis Z-scores accounting for sample overlapping**

Consider $L$ studies with overlapping samples, the $l^{th}$ study has sample size $n_l$ and Z-scores $\boldsymbol{Z_{l,score}}$. He et al. (He et al., 2022) propose a weighting scheme that accounts for sample overlapping and maximizes the power of meta-analysis. The optimal weights $w_l$ are the solutions of the following optimization problem

$$\text{minimize } \sum_{1\le i,j\le L} w_i w_j Cor.S_{ij}, \text{ subject to } \sum_l w_l\sqrt{n_l} = 1,\ w_l \ge 0,$$

where $Cor.S_{ij}$ denotes study correlations caused by sample overlapping and is calculated using corresponding Z-scores and LD.

**Logistic mixed effect model and score test**

For single variant association test (only one genetic variant is considered in the model), to test for the genotype effect $\beta$ with null hypothesis $H_0\colon \beta = 0$, we consider the following logistic mixed effect model:

$$\text{logit}(\pi_i) = \boldsymbol{X_i\alpha} + G_i\beta + b_i,$$

where $\pi_i = P(y_i = 1|\boldsymbol{X_i}, G_i, b_i)$ denotes the probability of the $i^{th}$ dichotomous phenotype, conditional on covariates $\boldsymbol{X_i}$, genotype $G_i$ and random effect $b_i$. The random effects are subject to a multivariate Gaussian distribution $\boldsymbol{b} \sim N(\boldsymbol{0}, \theta\boldsymbol{\Phi})$, where $\theta$ denotes variance component parameter and $\boldsymbol{\Phi}$ denotes the kinship matrix. With sample size $n$, the column vector $\boldsymbol{Y} = (Y_1, \cdots, Y_i, \cdots, Y_n)^T$ denotes phenotypes and the column vector $\boldsymbol{G} = (G_1, \cdots, G_i, \cdots, G_n)^T$ denotes the single variant genotypes.

The logistic mixed effect model can be rewritten as:

$$\xi_i = g(\pi_i) = g(\mu_i) = \boldsymbol{X_i^T\alpha} + G_i\beta + b_i,$$

with canonical link function $\xi_i = g(\pi_i) = \log\left(\frac{\pi_i}{1-\pi_i}\right)$. For dichotomous phenotype $y_i$, $\mu_i = \pi_i = P(y_i = 1|\boldsymbol{X_i}, G_i, b_i)$.

Given the random effect $\boldsymbol{b}$, the dichotomous phenotype $y_i$ is conditionally independent with mean and variance:

$$E(y_i|\boldsymbol{b}) = \mu_i$$

$$Var(y_i|\boldsymbol{b}) = \phi\mu_i(1-\mu_i),$$

where $\phi$ denotes the dispersion parameter and $\phi = 1$ for logistic mixed effect model.

For the $i^{th}$ individual, given random effects $\boldsymbol{b}$, the quasi-likelihood of fixed covariate effect $\boldsymbol{\alpha}$ and fixed genotype effect $\beta$ is:

$$ql_i(\boldsymbol{\alpha}, \beta; \boldsymbol{b}) = \int_{y_i}^{\mu_i} \frac{y_i - \mu}{\mu(1-\mu)} d\mu.$$

The log quasi-likelihood of $(\boldsymbol{\alpha}, \beta, \phi, \theta)$ integrated over the domain of random effect $\boldsymbol{b} \sim N(\boldsymbol{0}, \theta\boldsymbol{\Phi})$ is:

$$ql(\boldsymbol{\alpha}, \beta, \phi, \theta) = \log \int_{\boldsymbol{b}} \exp\left\{\sum_{i=1}^{n} ql_i(\boldsymbol{\alpha}, \beta; \boldsymbol{b})\right\} \times (2\pi)^{-\frac{n}{2}} |\theta\boldsymbol{\Phi}|^{-\frac{1}{2}} \times \exp\left\{-\frac{1}{2}\boldsymbol{b}^T(\theta\boldsymbol{\Phi})^{-1}\boldsymbol{b}\right\} d\boldsymbol{b}.$$

We define the function for simplicity:

$$f(\boldsymbol{b}) = \sum_{i=1}^{n} ql_i(\boldsymbol{\alpha}, \beta; \boldsymbol{b}) - \frac{1}{2}\boldsymbol{b}^T(\theta\boldsymbol{\Phi})^{-1}\boldsymbol{b}.$$

Using the Laplace's method to approximate the above function's integral, we have:

$$\int_{\boldsymbol{b}} \exp\{f(\boldsymbol{b})\}\, d\boldsymbol{b} \approx (2\pi)^{\frac{n}{2}} |-f''(\boldsymbol{b_0})|^{-\frac{1}{2}} \exp\{f(\boldsymbol{b_0})\},$$

where $\boldsymbol{b_0}$= $\text{argmax}_{\boldsymbol{b}} f(\boldsymbol{b})$ achieves $f(\boldsymbol{b})$'s global maximum at $\boldsymbol{b_0}$, and is the solution of $f'(\boldsymbol{b}) = 0$. In this case, the log quasi-likelihood is:

$$ql(\boldsymbol{\alpha}, \beta, \phi, \theta) \approx -\frac{1}{2}\log|\theta\boldsymbol{\Phi}| - \frac{1}{2}\log|-f''(\boldsymbol{b_0})| + f(\boldsymbol{b_0}).$$

To approximate the log quasi-likelihood $ql(\boldsymbol{\alpha}, \beta, \phi, \theta)$, we need to calculate the second partial derivative of $ql_i(\boldsymbol{\alpha}, \beta; \boldsymbol{b})$ with respect to random effects $\boldsymbol{b}$. The first partial derivative of $ql_i(\boldsymbol{\alpha}, \beta; \boldsymbol{b})$ with respect to $\boldsymbol{b}$ is:

$$\frac{\partial ql_i}{\partial \boldsymbol{b}} = \frac{\partial ql_i}{\partial \mu_i}\frac{\partial \mu_i}{\partial \xi_i}\frac{\partial \xi_i}{\partial \boldsymbol{b}} = \frac{y_i - \mu_i}{\mu_i(1-\mu_i)} \cdot \mu_i(1-\mu_i) \cdot \mathbb{I}_i^T = (y_i - \mu_i) \cdot \mathbb{I}_i^T,$$

where $g'(\mu_i) = \frac{1}{\mu_i(1-\mu_i)}$ and $\mathbb{I}_i$ is a $1 \times n$ row vector of indicators such that $b_i = \mathbb{I}_i\boldsymbol{b}$.

The second partial derivative of $ql_i(\boldsymbol{\alpha}, \beta; \boldsymbol{b})$ with respect to $\boldsymbol{b}$ is:

$$\frac{\partial^2 ql_i}{\partial \boldsymbol{b}\partial \boldsymbol{b}^T} = \frac{\partial(y_i - \mu_i)}{\partial \xi_i}\frac{\partial \xi_i}{\partial \boldsymbol{b}^T}\mathbb{I}_i^T + (y_i - \mu_i)\frac{\partial \mathbb{I}_i^T}{\partial \boldsymbol{b}^T} = -\mu_i(1-\mu_i)\mathbb{I}_i^T\mathbb{I}_i,$$

where the second term equals to 0 for canonical link $g(\mu_i) = \xi_i$ because $g'(\mu_i)v(\mu_i) = 1$. Therefore,

$$\begin{aligned}
ql(\boldsymbol{\alpha}, \beta, \phi, \theta) &\approx -\frac{1}{2}\log|\theta\boldsymbol{\Phi}| - \frac{1}{2}\log\left|\sum_{i=1}^{n}\mu_i(1-\mu_i)\mathbb{I}_i^T\mathbb{I}_i + (\theta\boldsymbol{\Phi})^{-1}\right| + \sum_{i=1}^{n} ql_i(\boldsymbol{\alpha}, \beta; \boldsymbol{b_0}) \\
&\quad - \frac{1}{2}\boldsymbol{b_0}^T(\theta\boldsymbol{\Phi})^{-1}\boldsymbol{b_0} \\
&= -\frac{1}{2}\log|\theta\boldsymbol{\Phi} \times (diag(\mu_i(1-\mu_i) + (\theta\boldsymbol{\Phi})^{-1})| + \sum_{i=1}^{n} ql_i(\boldsymbol{\alpha}, \beta; \boldsymbol{b_0}) - \frac{1}{2}\boldsymbol{b_0}^T(\theta\boldsymbol{\Phi})^{-1}\boldsymbol{b_0} \\
&= -\frac{1}{2}\log\left|\theta\boldsymbol{\Phi} \times diag\big(\mu_i(1-\mu_i)\big) + \boldsymbol{I}\right| + \sum_{i=1}^{n} ql_i(\boldsymbol{\alpha}, \beta; \boldsymbol{b_0}) - \frac{1}{2}\boldsymbol{b_0}^T(\theta\boldsymbol{\Phi})^{-1}\boldsymbol{b_0}.
\end{aligned}$$

We assume that the weight matrix $diag(\mu_i(1-\mu_i))$ changes slowly with respect to the conditional mean:

$$\frac{\partial diag(\mu_i(1-\mu_i))}{\partial \mu_i} \approx 0.$$

The partial derivatives of $ql(\boldsymbol{\alpha}, \beta, \phi, \theta)$ are:

$$\frac{\partial ql(\boldsymbol{\alpha}, \beta, \phi, \theta)}{\partial \boldsymbol{\alpha}} = \sum_{i=1}^{n} \frac{\partial ql_i}{\partial \mu_i}\frac{\partial \mu_i}{\partial \xi_i}\frac{\partial \xi_i}{\partial \boldsymbol{\alpha}} = \sum_{i=1}^{n} \frac{y_i - \mu_i}{\mu_i(1-\mu_i)} \cdot \mu_i(1-\mu_i) \cdot \boldsymbol{X}_{\boldsymbol{i}}^T = \boldsymbol{X}^T(\boldsymbol{Y}-\boldsymbol{\mu})$$

$$\frac{\partial ql(\boldsymbol{\alpha}, \beta, \phi, \theta)}{\partial \beta} = \sum_{i=1}^{n} \frac{\partial ql_i}{\partial \mu_i}\frac{\partial \mu_i}{\partial \xi_i}\frac{\partial \xi_i}{\partial \beta} = \sum_{i=1}^{n} \frac{y_i - \mu_i}{\mu_i(1-\mu_i)} \cdot \mu_i(1-\mu_i) \cdot G_i = \boldsymbol{G}^T(\boldsymbol{Y}-\boldsymbol{\mu})$$

$$\frac{\partial ql(\boldsymbol{\alpha}, \beta, \phi, \theta)}{\partial \boldsymbol{b}} = \sum_{i=1}^{n} \frac{\partial ql_i}{\partial \mu_i}\frac{\partial \mu_i}{\partial \xi_i}\frac{\partial \xi_i}{\partial \boldsymbol{b}} - (\theta\boldsymbol{\Phi})^{-1}\boldsymbol{b} = \sum_{i=1}^{n} (y_i - \mu_i) \cdot \mathbb{I}_i^T - (\theta\boldsymbol{\Phi})^{-1}\boldsymbol{b} = (\boldsymbol{Y}-\boldsymbol{\mu}) - (\theta\boldsymbol{\Phi})^{-1}\boldsymbol{b}.$$

Under the null hypothesis $H_0: \beta = 0$, and estimates of dispersion parameter $\phi$, kinship matrix $\boldsymbol{\Phi}$ and variance component parameter $\theta$ are known, the fixed covariate effect $\widehat{\boldsymbol{\alpha}}(\phi, \theta)$ and random effects $\widehat{\boldsymbol{b}}(\phi, \theta, \boldsymbol{\Phi})$ are estimated simultaneously by maximizing the log quasi-likelihood $ql(\boldsymbol{\alpha}, \beta, \phi, \theta)$. Therefore, $\widehat{\boldsymbol{b}}(\phi, \theta, \boldsymbol{\Phi}) = \boldsymbol{b_0}(\widehat{\boldsymbol{\alpha}}(\phi, \theta), \beta = 0)$ because at $\boldsymbol{b_0}$, $ql(\boldsymbol{\alpha}, \beta, \phi, \theta)$ achieves global maximum.

The working vector of dichotomous phenotypes is $\widetilde{\boldsymbol{Y}} = \left(\widetilde{Y}_1, \cdots, \widetilde{Y}_n\right)^T$, where $\widetilde{\boldsymbol{Y}} - \boldsymbol{\xi} = g'(\boldsymbol{\mu})(\boldsymbol{Y}-\boldsymbol{\mu})$. The solution $(\boldsymbol{\alpha}, \beta)$ of linear system under the null hypothesis is:

$$\begin{cases} \boldsymbol{X}^T(\boldsymbol{Y}-\boldsymbol{\mu}) = 0 \\ (\boldsymbol{Y}-\boldsymbol{\mu}) - (\theta\boldsymbol{\Phi})^{-1}\boldsymbol{b} = \boldsymbol{0} \end{cases}$$

which is equivalent to solving the following equation:

$$\begin{bmatrix} \boldsymbol{X}^T diag(\boldsymbol{\mu}(\boldsymbol{1}-\boldsymbol{\mu}))\boldsymbol{X} & \boldsymbol{X}^T diag(\boldsymbol{\mu}(\boldsymbol{1}-\boldsymbol{\mu})) \\ diag(\boldsymbol{\mu}(\boldsymbol{1}-\boldsymbol{\mu}))\boldsymbol{X} & (\theta\boldsymbol{\Phi})^{-1} + diag(\boldsymbol{\mu}(\boldsymbol{1}-\boldsymbol{\mu})) \end{bmatrix} \begin{bmatrix} \boldsymbol{\alpha} \\ \boldsymbol{b} \end{bmatrix} = \begin{bmatrix} \boldsymbol{X}^T diag(\boldsymbol{\mu}(\boldsymbol{1}-\boldsymbol{\mu}))\widetilde{\boldsymbol{Y}} \\ diag(\boldsymbol{\mu}(\boldsymbol{1}-\boldsymbol{\mu}))\widetilde{\boldsymbol{Y}} \end{bmatrix}$$

Let $\boldsymbol{\Omega} = diag\left(\frac{\boldsymbol{1}}{\boldsymbol{\mu}(\boldsymbol{1}-\boldsymbol{\mu})}\right) + \theta\boldsymbol{\Phi}$ and $\boldsymbol{\Psi} = \boldsymbol{\Omega}^{-\boldsymbol{1}} - \boldsymbol{\Omega}^{-\boldsymbol{1}}\boldsymbol{X}(\boldsymbol{X}^T\boldsymbol{\Omega}^{-\boldsymbol{1}}\boldsymbol{X})^{-\boldsymbol{1}}\boldsymbol{X}^T\boldsymbol{\Omega}^{-\boldsymbol{1}}$ (projection matrix). The solution to maximize the log quasi-likelihood is:

$$\begin{cases} \widehat{\boldsymbol{\alpha}} = (\boldsymbol{X}^T\boldsymbol{\Omega}^{-\boldsymbol{1}}\boldsymbol{X})^{-\boldsymbol{1}}\boldsymbol{X}^T\boldsymbol{\Omega}^{-\boldsymbol{1}}\widetilde{\boldsymbol{Y}} \\ \widehat{\boldsymbol{b}} = (\theta\boldsymbol{\Phi})\boldsymbol{\Omega}^{-\boldsymbol{1}}(\widetilde{\boldsymbol{Y}} - \boldsymbol{X}\widehat{\boldsymbol{\alpha}}) \end{cases}$$

$$\widetilde{\boldsymbol{Y}} - \widehat{\boldsymbol{\xi}} = \widetilde{\boldsymbol{Y}} - \boldsymbol{X}\widehat{\boldsymbol{\alpha}} - \widehat{\boldsymbol{b}} = (\boldsymbol{I} - \theta\boldsymbol{\Phi}\boldsymbol{\Omega}^{-\boldsymbol{1}})(\widetilde{\boldsymbol{Y}} - \boldsymbol{X}\widehat{\boldsymbol{\alpha}}) = diag\left(\frac{\boldsymbol{1}}{\boldsymbol{\mu}(\boldsymbol{1}-\boldsymbol{\mu})}\right)\boldsymbol{\Omega}^{-\boldsymbol{1}}(\widetilde{\boldsymbol{Y}} - \boldsymbol{X}\widehat{\boldsymbol{\alpha}})$$
$$= diag\left(\frac{\boldsymbol{1}}{\boldsymbol{\mu}(\boldsymbol{1}-\boldsymbol{\mu})}\right)\boldsymbol{\Psi}\widetilde{\boldsymbol{Y}},$$

Under the null hypothesis $H_0: \beta = 0$ and estimates of parameters $(\widehat{\boldsymbol{\alpha}}, \widehat{\phi}, \widehat{\theta})$, the score test is constructed by plugging in $(\widehat{\boldsymbol{\alpha}}, \widehat{\phi}, \widehat{\theta}, \beta = 0)$ in $\frac{\partial ql(\boldsymbol{\alpha}, \beta, \phi, \theta)}{\partial \beta}$ as follows:

$$T = \frac{\partial ql(\widehat{\boldsymbol{\alpha}}, \widehat{\phi}, \widehat{\theta}, \beta = 0)}{\partial \beta} = \boldsymbol{G}^T(\boldsymbol{Y} - \widehat{\boldsymbol{\mu}}) = \boldsymbol{G}^T(\widetilde{\boldsymbol{Y}} - \widehat{\boldsymbol{\xi}})\frac{1}{g'(\widehat{\boldsymbol{\mu}})} = \boldsymbol{G}^T diag\left(\frac{\boldsymbol{1}}{\widehat{\boldsymbol{\mu}}(\boldsymbol{1}-\widehat{\boldsymbol{\mu}})}\right)\widehat{\boldsymbol{\Psi}}\widetilde{\boldsymbol{Y}} diag(\widehat{\boldsymbol{\mu}}(\boldsymbol{1}-\widehat{\boldsymbol{\mu}}))$$
$$= \boldsymbol{G}^T\widehat{\boldsymbol{\Psi}}\widetilde{\boldsymbol{Y}},$$

with corresponding variance being:

$$\mathrm{Var}(T|H_0) = E\left\{\frac{\partial ql(\widehat{\boldsymbol{\alpha}}, \widehat{\phi}, \widehat{\theta}, \beta = 0)}{\partial \beta}\frac{\partial ql(\widehat{\boldsymbol{\alpha}}, \widehat{\phi}, \widehat{\theta}, \beta = 0)}{\partial \beta^T}\right\} = E\left\{\boldsymbol{G}^T\widehat{\boldsymbol{\Psi}}\widetilde{\boldsymbol{Y}}\left(\boldsymbol{G}^T\widehat{\boldsymbol{\Psi}}\widetilde{\boldsymbol{Y}}\right)^T\right\} = \boldsymbol{G}^T\widehat{\boldsymbol{\Psi}}\boldsymbol{G}.$$

**Linear mixed effect model and score test**

For single variant association test (only one genetic variant is considered in the model), to test for the genotype effect $\beta$ with null hypothesis $H_0$: $\beta = 0$, we consider the following linear mixed effect model:

$$y_i = \boldsymbol{X_i\alpha} + G_i\beta + b_i + \epsilon_i,$$

which can be rewritten as:

$$\xi_i = \mu_i = \mathrm{E}(y_i) = \boldsymbol{X_i^T\alpha} + G_i\beta + b_i.$$

with canonical link function $\xi_i = g(\mu_i) = \mu_i$.

For the $i^{th}$ individual, given random effects $\boldsymbol{b}$, the quasi-likelihood of fixed covariate effect $\boldsymbol{\alpha}$ and fixed genotype effect $\beta$ is:

$$ql_i(\boldsymbol{\alpha}, \beta; \boldsymbol{b}) = \int_{y_i}^{\mu_i} \frac{y_i - \mu}{\phi} d\mu,$$

where $\phi$ denotes the dispersion parameter and $\phi = \sigma^2$ for linear mixed model.

The log quasi-likelihood of $(\boldsymbol{\alpha}, \beta, \phi, \theta)$ integrated over the domain of random effects $\boldsymbol{b} \sim N(\boldsymbol{0}, \theta\boldsymbol{\Phi})$ is:

$$ql(\boldsymbol{\alpha}, \beta, \phi, \theta) = \log \int_{\boldsymbol{b}} \exp\left\{\sum_{i=1}^{n} ql_i(\boldsymbol{\alpha}, \beta; \boldsymbol{b})\right\} \times (2\pi)^{-\frac{n}{2}} |\theta\boldsymbol{\Phi}|^{-\frac{1}{2}} \times \exp\left\{-\frac{1}{2}\boldsymbol{b}^T(\theta\boldsymbol{\Phi})^{-1}\boldsymbol{b}\right\} d\boldsymbol{b}.$$

where θ is the variance component parameter, and $\boldsymbol{\Phi}$ is the kinship matrix.

We define the function for simplicity:

$$f(\boldsymbol{b}) = \sum_{i=1}^{n} ql_i(\boldsymbol{\alpha}, \beta; \boldsymbol{b}) - \frac{1}{2}\boldsymbol{b}^T(\theta\boldsymbol{\Phi})^{-1}\boldsymbol{b}.$$

Using the Laplace's method to approximate the above function's integral, we have:

$$\int_{\boldsymbol{b}} \exp\{f(\boldsymbol{b})\}\, d\boldsymbol{b} \approx (2\pi)^{\frac{n}{2}} |-f''(\boldsymbol{b_0})|^{-\frac{1}{2}} \exp\{f(\boldsymbol{b_0})\},$$

where $\boldsymbol{b_0}$= $\mathrm{argmax}_{\boldsymbol{b}} f(\boldsymbol{b})$ achieves $f(\boldsymbol{b})$'s global maximum at $\boldsymbol{b_0}$, and is the solution of $f'(\boldsymbol{b}) = 0$. In this case, the log quasi-likelihood is:

$$ql(\boldsymbol{\alpha}, \beta, \phi, \theta) \approx -\frac{1}{2} log|\theta\boldsymbol{\Phi}| - \frac{1}{2} log|-f''(\boldsymbol{b_0})| + f(\boldsymbol{b_0}).$$

To approximate the log quasi-likelihood $ql(\boldsymbol{\alpha}, \beta, \phi, \theta)$, we need to calculate the second partial derivative of $ql_i(\boldsymbol{\alpha}, \beta; \boldsymbol{b})$ with respect to random effects $\boldsymbol{b}$. The first partial derivative of $ql_i(\boldsymbol{\alpha}, \beta; \boldsymbol{b})$ with respect to $\boldsymbol{b}$ is:

$$\frac{\partial ql_i}{\partial \boldsymbol{b}} = \frac{\partial ql_i}{\partial \mu_i}\frac{\partial \mu_i}{\partial \xi_i}\frac{\partial \xi_i}{\partial \boldsymbol{b}} = \frac{y_i - \mu_i}{\phi} \cdot 1 \cdot \mathbb{I}_i^T,$$

where $\mathbb{I}_i$ is a $1 \times n$ row vector of indicators such that $b_i = \mathbb{I}_i\boldsymbol{b}$.

The second partial derivative of $ql_i(\boldsymbol{\alpha}, \beta; \boldsymbol{b})$ with respect to $\boldsymbol{b}$ is:

$$\frac{\partial^2 ql_i}{\partial \boldsymbol{b}\partial \boldsymbol{b}^T} = \frac{\partial \frac{y_i - \mu_i}{\phi}}{\partial \xi_i}\frac{\partial \xi_i}{\partial \boldsymbol{b}^T}\mathbb{I}_i^T + \frac{y_i - \mu_i}{\phi}\frac{\partial \mathbb{I}_i^T}{\partial \boldsymbol{b}^T} = -\frac{1}{\phi}\mathbb{I}_i^T\mathbb{I}_i,$$

where the second term equals to 0 for canonical link $g(\mu_i) = \xi_i$.

$$ql(\boldsymbol{\alpha}, \beta, \phi, \theta) \approx -\frac{1}{2} log|\theta\boldsymbol{\Phi}| - \frac{1}{2} log \left|\sum_{i=1}^{n} \frac{1}{\phi} \mathbb{I}_i^T \mathbb{I}_i + (\theta\boldsymbol{\Phi})^{-1}\right| + \sum_{i=1}^{n} ql_i(\boldsymbol{\alpha}, \beta; \boldsymbol{b_0}) - \frac{1}{2} \boldsymbol{b_0^T} (\theta\boldsymbol{\Phi})^{-1} \boldsymbol{b_0}$$

$$= -\frac{1}{2} log \left|\theta\boldsymbol{\Phi} \times \left(diag\left(\frac{1}{\phi}\right) + (\theta\boldsymbol{\Phi})^{-1}\right)\right| + \sum_{i=1}^{n} ql_i(\boldsymbol{\alpha}, \beta; \boldsymbol{b_0}) - \frac{1}{2} \boldsymbol{b_0^T} (\theta\boldsymbol{\Phi})^{-1} \boldsymbol{b_0}$$

$$= -\frac{1}{2} log \left|diag\left(\frac{\theta\boldsymbol{\Phi}}{\phi}\right) + \boldsymbol{I}\right| + \sum_{i=1}^{n} ql_i(\boldsymbol{\alpha}, \beta; \boldsymbol{b_0}) - \frac{1}{2} \boldsymbol{b_0^T} (\theta\boldsymbol{\Phi})^{-1} \boldsymbol{b_0}.$$

The dispersion parameter $\phi = \sigma^2$ is a constant with respect to mean $\mu_i$ and hence

$$\frac{\partial \phi}{\partial \mu_i} \equiv 0.$$

The partial derivatives of $ql(\boldsymbol{\alpha}, \beta, \phi, \theta)$ are:

$$\frac{\partial ql(\boldsymbol{\alpha}, \beta, \phi, \theta)}{\partial \boldsymbol{\alpha}} = \sum_{i=1}^{n} \frac{\partial ql_i}{\partial \mu_i} \frac{\partial \mu_i}{\partial \xi_i} \frac{\partial \xi_i}{\partial \boldsymbol{\alpha}} = \sum_{i=1}^{n} \frac{y_i - \mu_i}{\phi} \cdot 1 \cdot \boldsymbol{X_i^T} = \boldsymbol{X}^T \text{diag}(\frac{1}{\phi})(\boldsymbol{Y} - \boldsymbol{\mu})$$

$$\frac{\partial ql(\boldsymbol{\alpha}, \beta, \phi, \theta)}{\partial \beta} = \sum_{i=1}^{n} \frac{\partial ql_i}{\partial \mu_i} \frac{\partial \mu_i}{\partial \xi_i} \frac{\partial \xi_i}{\partial \beta} = \sum_{i=1}^{n} \frac{y_i - \mu_i}{\phi} \cdot 1 \cdot G_i = \boldsymbol{G}^T \text{diag}(\frac{1}{\phi})(\boldsymbol{Y} - \boldsymbol{\mu})$$

$$\frac{\partial ql(\boldsymbol{\alpha}, \beta, \phi, \theta)}{\partial \boldsymbol{b}} = \sum_{i=1}^{n} \frac{\partial ql_i}{\partial \mu_i} \frac{\partial \mu_i}{\partial \xi_i} \frac{\partial \xi_i}{\partial \boldsymbol{b}} - (\theta\boldsymbol{\Phi})^{-1}\boldsymbol{b} = \sum_{i=1}^{n} \frac{y_i - \mu_i}{\phi} \mathbb{I}_i^T - (\theta\boldsymbol{\Phi})^{-1}\boldsymbol{b}$$

$$= \text{diag}\left(\frac{1}{\phi}\right)(\boldsymbol{Y} - \boldsymbol{\mu}) - (\theta\boldsymbol{\Phi})^{-1}\boldsymbol{b}.$$

Under the null hypothesis $H_0: \beta = 0$, and estimates of dispersion parameter $\phi$, kinship matrix $\boldsymbol{\Phi}$ and variance component parameter $\theta$ are known, the fixed covariate effect $\widehat{\boldsymbol{\alpha}}(\phi, \theta)$ and random effect $\widehat{\boldsymbol{b}}(\phi, \theta, \boldsymbol{\Phi})$ are estimated simultaneously by maximizing the log quasi-likelihood $ql(\boldsymbol{\alpha}, \beta, \phi, \theta)$. Therefore, $\widehat{\boldsymbol{b}}(\phi, \theta, \boldsymbol{\Phi}) = \boldsymbol{b_0}(\widehat{\boldsymbol{\alpha}}(\phi, \theta), \beta = 0)$ because at $\boldsymbol{b_0}$, $ql(\boldsymbol{\alpha}, \beta, \phi, \theta)$ achieves global maximum.

The working vector of quantitative phenotypes are $\widetilde{\boldsymbol{Y}} = \left(\tilde{Y}_1, \cdots, \tilde{Y}_n\right)^T$, where $\widetilde{\boldsymbol{Y}} - \boldsymbol{\xi} = \boldsymbol{Y} - \boldsymbol{\mu}$. The solution $(\boldsymbol{\alpha}, \beta)$ of linear system under the null hypothesis is:

$$\begin{cases} \boldsymbol{X}^T \text{diag}(\frac{1}{\phi})(\boldsymbol{Y} - \boldsymbol{\mu}) = 0 \\ \text{diag}(\frac{1}{\phi})(\boldsymbol{Y} - \boldsymbol{\mu}) - (\theta\boldsymbol{\Phi})^{-1}\boldsymbol{b} = \boldsymbol{0} \end{cases}$$

which is equivalent to solving equation:

$$\begin{bmatrix} \boldsymbol{X}^T \text{diag}(\frac{1}{\phi})\boldsymbol{X} & \boldsymbol{X}^T \text{diag}(\frac{1}{\phi}) \\ \text{diag}(\frac{1}{\phi})\boldsymbol{X} & (\theta\boldsymbol{\Phi})^{-1} + \text{diag}(\frac{1}{\phi}) \end{bmatrix} \begin{bmatrix} \boldsymbol{\alpha} \\ \boldsymbol{b} \end{bmatrix} = \begin{bmatrix} \boldsymbol{X}^T \text{diag}(\frac{1}{\phi})\widetilde{\boldsymbol{Y}} \\ \text{diag}(\frac{1}{\phi})\widetilde{\boldsymbol{Y}} \end{bmatrix}$$

Let $\boldsymbol{\Omega} = \text{diag}(\phi) + \theta\boldsymbol{\Phi}$ and $\boldsymbol{\Psi} = \boldsymbol{\Omega^{-1}} - \boldsymbol{\Omega^{-1}X}(\boldsymbol{X^T\Omega^{-1}X})^{-1}\boldsymbol{X^T\Omega^{-1}}$. The solution to maximize the log quasi-likelihood is:

$$\begin{cases} \widehat{\boldsymbol{\alpha}} = (\boldsymbol{X}^T\boldsymbol{\Omega}^{-1}\boldsymbol{X})^{-1}\boldsymbol{X}^T\boldsymbol{\Omega}^{-1}\widetilde{\boldsymbol{Y}} \\ \widehat{\boldsymbol{b}} = \theta\boldsymbol{\Phi}\boldsymbol{\Omega}^{-1}(\widetilde{\boldsymbol{Y}} - \boldsymbol{X}\widehat{\boldsymbol{\alpha}}) \end{cases}$$

$$\widetilde{\boldsymbol{Y}} - \widehat{\boldsymbol{\xi}} = \widetilde{\boldsymbol{Y}} - \boldsymbol{X}\widehat{\boldsymbol{\alpha}} - \widehat{\boldsymbol{b}} = (\boldsymbol{I} - \theta\boldsymbol{\Phi}\boldsymbol{\Omega}^{-1})(\widetilde{\boldsymbol{Y}} - \boldsymbol{X}\widehat{\boldsymbol{\alpha}}) = \boldsymbol{diag}(\phi)\boldsymbol{\Omega}^{-1}(\widetilde{\boldsymbol{Y}} - \boldsymbol{X}\widehat{\boldsymbol{\alpha}}) = \boldsymbol{diag}(\phi)\boldsymbol{\Psi}\widetilde{\boldsymbol{Y}}.$$

Under the null hypothesis $H_0$: $\beta = 0$ and estimates of parameters $(\widehat{\boldsymbol{\alpha}}, \hat{\phi}, \hat{\theta})$, the score test is constructed by plugging in $(\widehat{\boldsymbol{\alpha}}, \hat{\phi}, \hat{\theta}, \beta = 0)$ in $\frac{\partial ql(\boldsymbol{\alpha},\beta,\phi,\theta)}{\partial\beta}$ as follows:

$$T = \frac{\partial ql(\widehat{\boldsymbol{\alpha}}, \hat{\phi}, \hat{\theta}, \beta = 0)}{\partial\beta} = \boldsymbol{G}^T \text{diag}\left(\frac{1}{\hat{\phi}}\right)(\boldsymbol{Y} - \widehat{\boldsymbol{\mu}}) = \boldsymbol{G}^T \text{diag}\left(\frac{1}{\hat{\phi}}\right)(\widetilde{\boldsymbol{Y}} - \widehat{\boldsymbol{\xi}}) = \boldsymbol{G}^T \text{diag}\left(\frac{1}{\hat{\phi}}\right)\boldsymbol{diag}(\hat{\phi})\widehat{\boldsymbol{\Psi}}\widetilde{\boldsymbol{Y}}$$
$$= \boldsymbol{G}^T\widehat{\boldsymbol{\Psi}}\widetilde{\boldsymbol{Y}},$$

with corresponding variance being:

$$\text{Var}(T|H_0) = E\left\{\frac{\partial ql(\widehat{\boldsymbol{\alpha}}, \hat{\phi}, \hat{\theta}, \beta = 0)}{\partial\beta}\frac{\partial ql(\widehat{\boldsymbol{\alpha}}, \hat{\phi}, \hat{\theta}, \beta = 0)}{\partial\beta^T}\right\} = \boldsymbol{G}^T\widehat{\boldsymbol{\Psi}}\widetilde{\boldsymbol{Y}}(\boldsymbol{G}^T\widehat{\boldsymbol{\Psi}}\widetilde{\boldsymbol{Y}})^T = \boldsymbol{G}^T\widehat{\boldsymbol{\Psi}}\boldsymbol{G}$$

$$\widetilde{\boldsymbol{Y}}\widetilde{\boldsymbol{Y}}^{\boldsymbol{T}} = (\widetilde{\boldsymbol{Y}} - \widehat{\boldsymbol{\xi}})(\widetilde{\boldsymbol{Y}} - \widehat{\boldsymbol{\xi}})^{\boldsymbol{T}} = (\boldsymbol{Y} - \widehat{\boldsymbol{\mu}})(\boldsymbol{Y} - \widehat{\boldsymbol{\mu}})^T = \boldsymbol{Y}\boldsymbol{Y}^T.$$

## Supplementary Figures

### Supplementary Figure 1: Three-generation pedigree

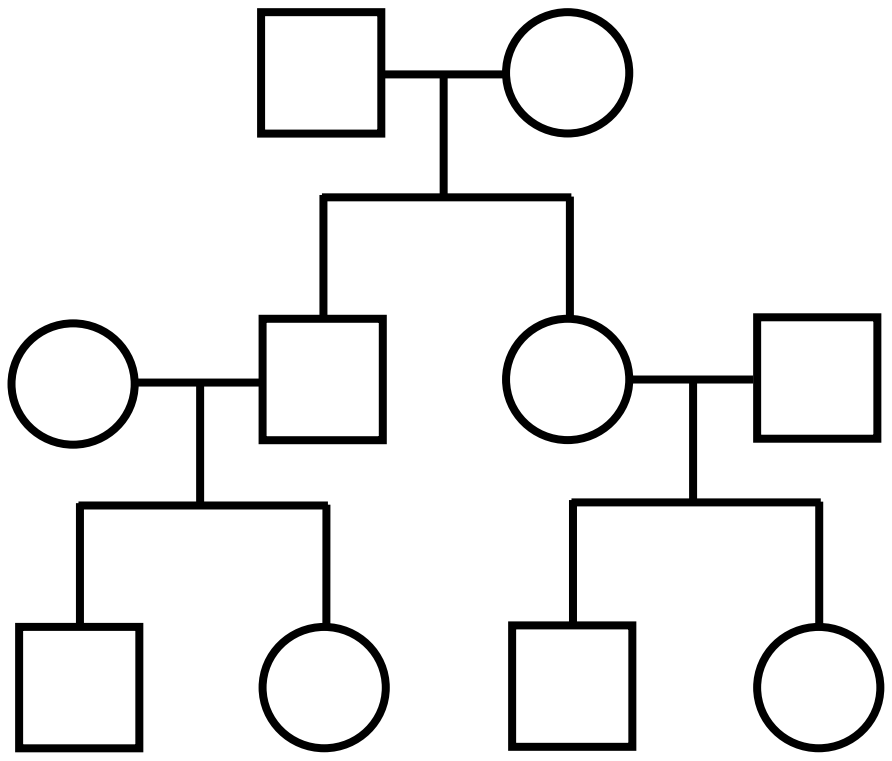

**Supplementary Figure 2: Empirical simulation studies of GhostKnockoffs for different levels of sample relatedness at FDR level 0.1.** We present FDR and power estimation of GhostKnockoffs and the second-order knockoffs based on 1000 replicates of 10,000 samples for quantitative and dichotomous phenotypes. Dichotomous phenotypes are of varying levels of relatedness due to different sampling schemes of case control family studies. GhostKnockoff, mixed model score test/score test: multiple GhostKnockoffs based on Z-scores converted from p-values of mixed effect model score test by adjusting kinship among phenotypes/score test without phenotype kinship adjustment. IndividualData knockoff, mixed model score test/score test: multiple second-order knockoffs based on Z-scores converted from p-values of mixed effect model score test by adjusting kinship among phenotypes/score test without phenotype kinship adjustment.

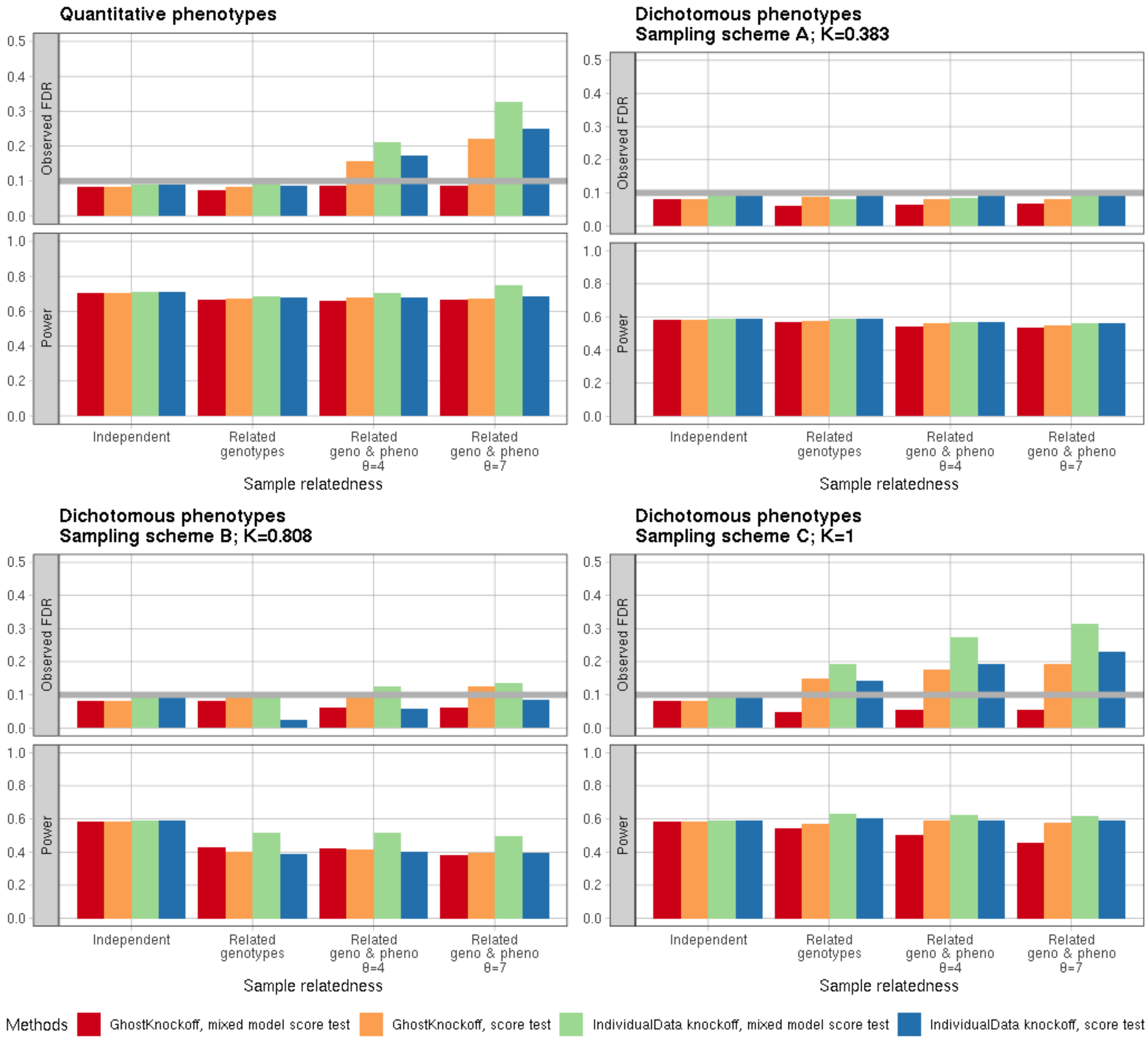

**Supplementary Figure 3: Empirical simulation studies of GhostKnockoffs for different levels of sample relatedness at FDR level 0.2.** We present FDR and power estimation of GhostKnockoffs and the second-order knockoffs based on 1000 replicates of 10,000 samples for quantitative and dichotomous phenotypes. Dichotomous phenotypes are of varying levels of relatedness due to different sampling schemes of case control family studies. GhostKnockoff, mixed model score test/score test: multiple GhostKnockoffs based on Z-scores converted from p-values of mixed effect model score test by adjusting kinship among phenotypes/score test without phenotype kinship adjustment. IndividualData knockoff, mixed model score test/score test: multiple second-order knockoffs based on Z-scores converted from p-values of mixed effect model score test by adjusting kinship among phenotypes/score test without phenotype kinship adjustment.

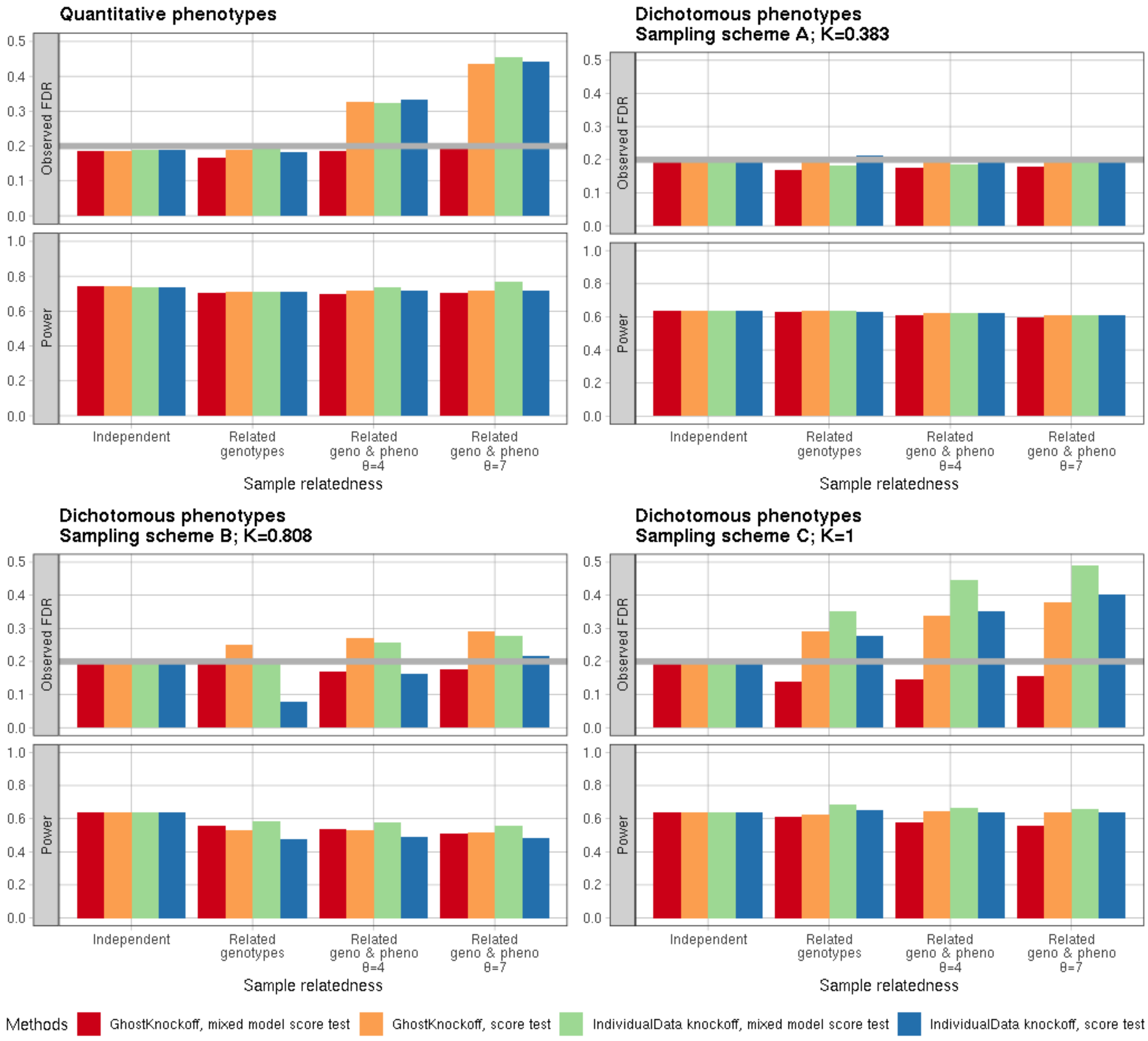

**Supplementary Figure 4: Empirical simulation studies of GhostKnockoffs for quantitative phenotypes.** We present FDR and power estimation of GhostKnockoffs and the second-order knockoffs based on 1000 replicates of 10,000 samples for quantitative phenotypes. The top panel shows FDR and power estimates of two methods for independent genotypes and phenotypes. The bottom panels show FDR and power estimates of two methods for related genotypes (simulated using gene dropping algorithm) and phenotypes (simulated from mixed effect model with different values of random effect's variance component parameter $\theta$). GhostKnockoff, mixed model score test/score test: multiple GhostKnockoffs based on Z-scores converted from p-values of mixed effect model score test by adjusting kinship among phenotypes/score test without phenotype kinship adjustment. IndividualData knockoff, mixed model score test/score test: multiple second-order knockoffs based on Z-scores converted from p-values of mixed effect model score test by adjusting kinship among phenotypes/score test without phenotype kinship adjustment.

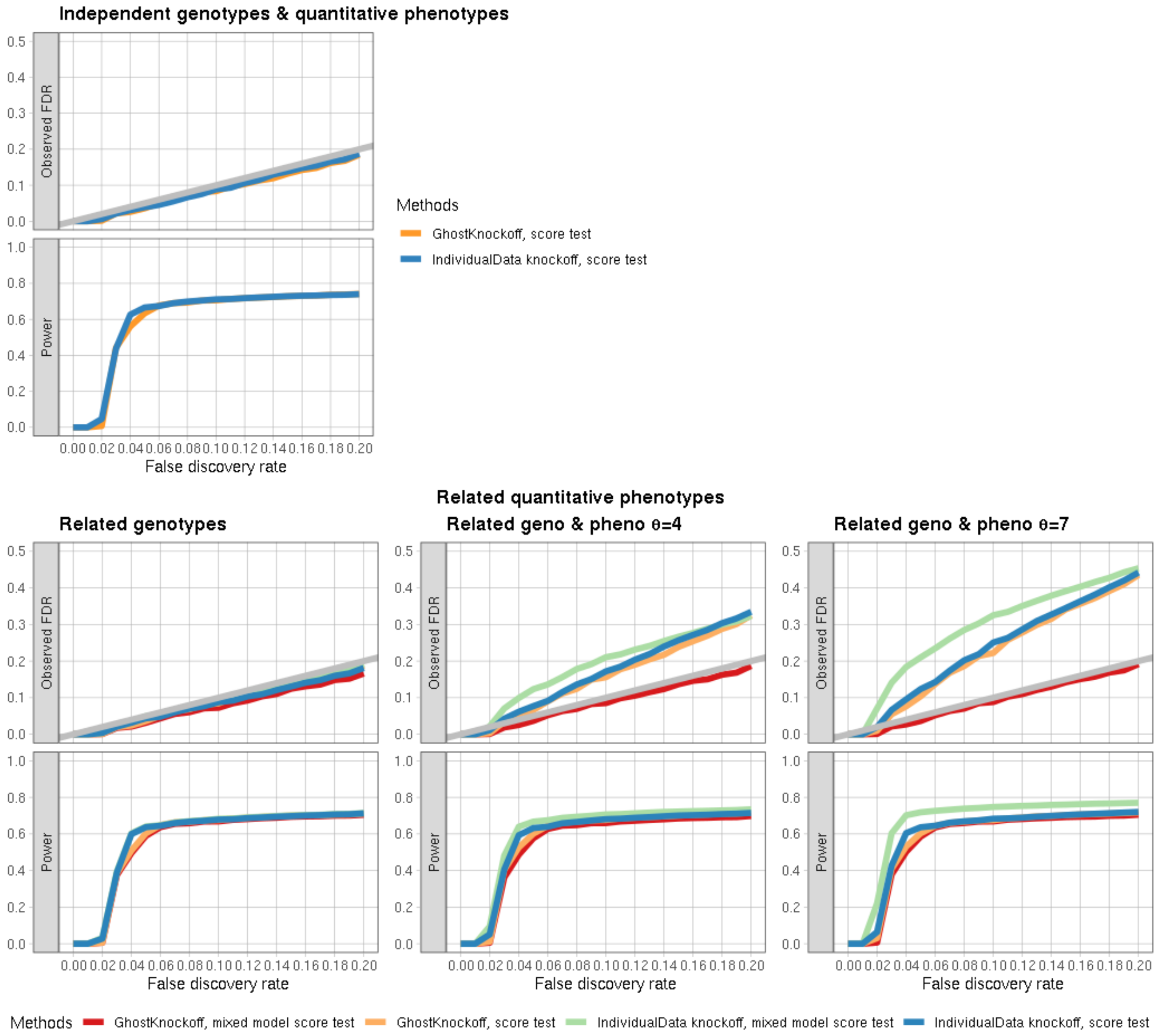

**Supplementary Figure 5: Empirical simulation studies of GhostKnockoffs for dichotomous phenotypes.** We present FDR and power estimation of GhostKnockoffs and the second-order knockoffs based on 1000 replicates of 10,000 samples for dichotomous phenotypes based on different sampling schemes of case control family studies. **Scheme A.** We randomly select 5000 cases and 5000 controls from the simulated foundation data. **Scheme B.** We randomly select 500 case families (number of cases greater than or equal to 1 per family) and 500 control families (0 cases per family) from the simulated foundation data. **Scheme C**. We select 5000 cases by including all cases from case families and randomly select 5000 controls from all control families based on the simulated foundation data. The top panel shows FDR and power estimates of two methods for independent genotypes and phenotypes. The bottom panels show FDR and power estimates of two methods for related genotypes (simulated using gene dropping algorithm) and phenotypes (simulated from mixed effect model with different values of random effect's variance component parameter $\theta$). GhostKnockoff, mixed model score test/score test: multiple GhostKnockoffs based on Z-scores converted from p-values of mixed effect model score test by adjusting kinship among phenotypes/score test without phenotype kinship adjustment. IndividualData knockoff, mixed model score test/score test: multiple second-order knockoffs based on Z-scores converted from p-values of mixed effect model score test by adjusting kinship among phenotypes/ score test without phenotype kinship adjustment.

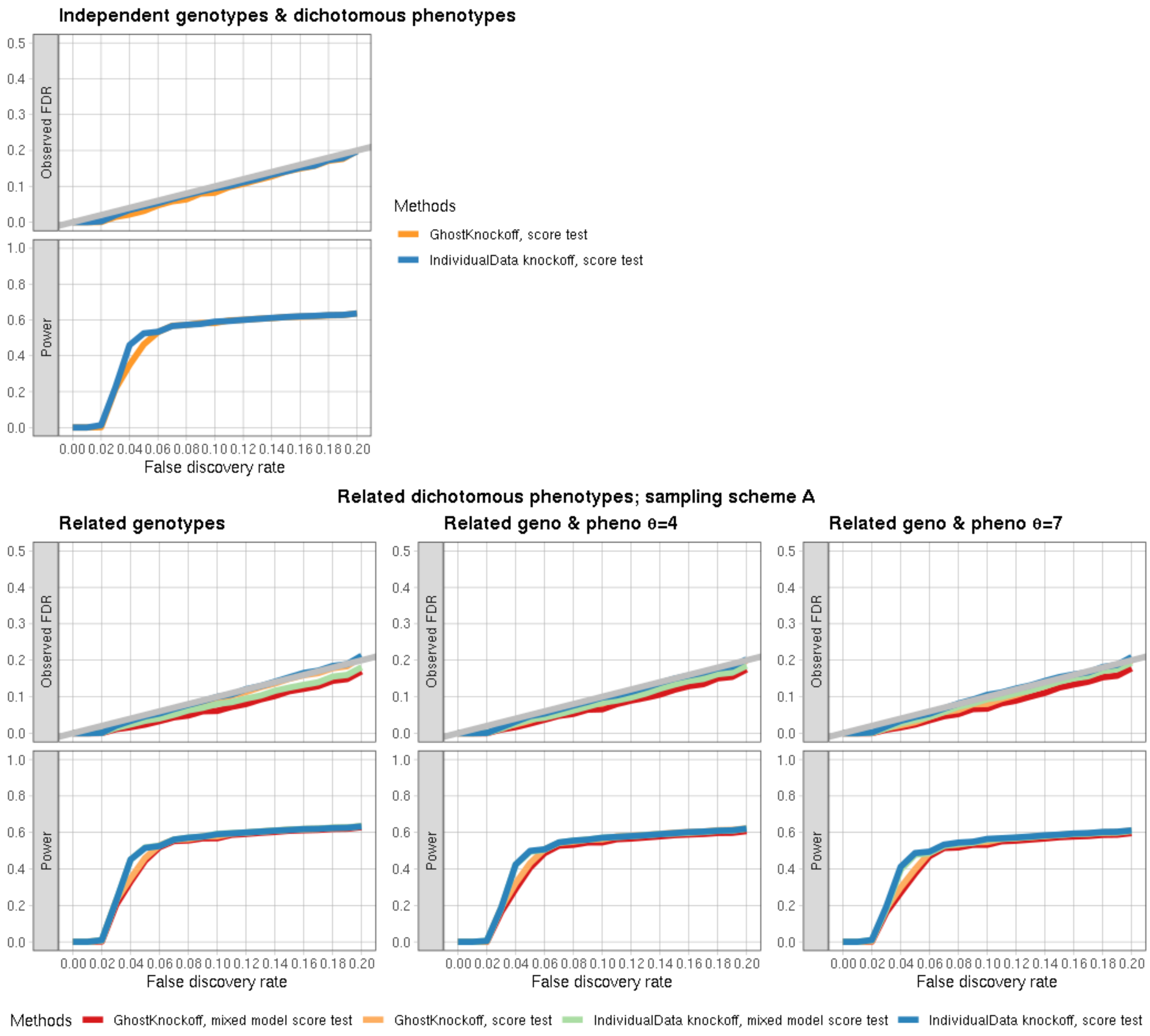

**Supplementary Figure 5 (continue): Empirical simulation studies of GhostKnockoffs for dichotomous phenotypes.**

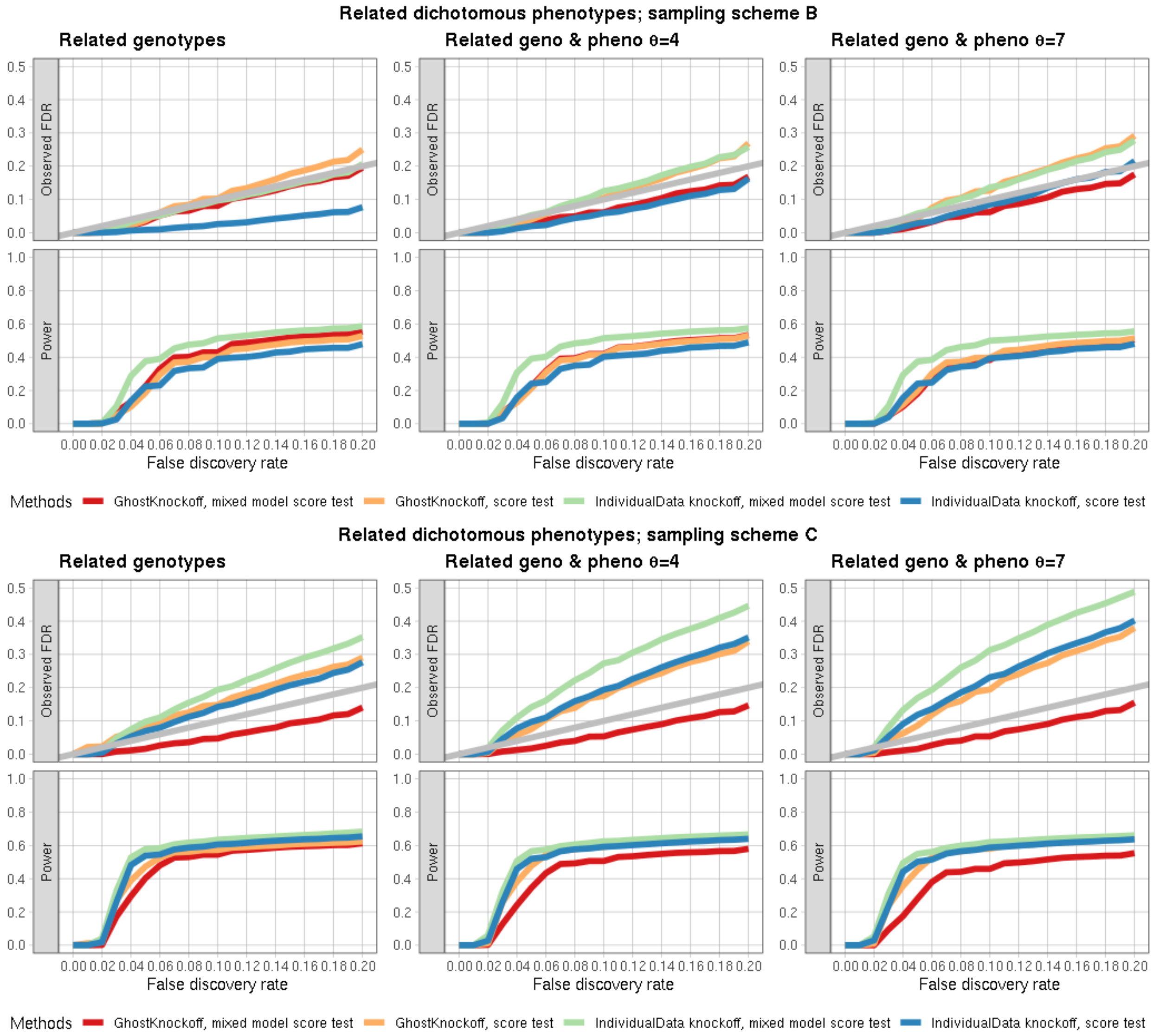

**Supplementary Figure 6: QQ plots of empirical simulation studies.** We present QQ plots of empirical simulation studies based on 1000 replicates of 10,000 samples for quantitative and dichotomous phenotypes. The left column panel shows QQ plots for independent/related genotypes and quantitative phenotypes. The right three right column panels show QQ plots for independent/related genotypes and dichotomous phenotypes simulated from three sampling schemes. **Scheme A.** We randomly select 5000 cases and 5000 controls from the simulated foundation data. **Scheme B.** We randomly select 500 case families (number of cases greater than or equal to 1 per family) and 500 control families (0 cases per family) from the simulated foundation data. **Scheme C**. We select 5000 cases by including all cases from case families and randomly select 5000 controls from all control families based on the simulated foundation data. The genomic inflation factors, denoted as “lambda gc”, of two association tests (score test and mixed model score test) are listed in the upper-left corner of each QQ plot. The divergence between score test and mixed model score test demonstrates the phenotype relatedness due to (1) random effects of mixed effect model to simulate phenotypes, (2) additional relatedness caused by sampling schemes.

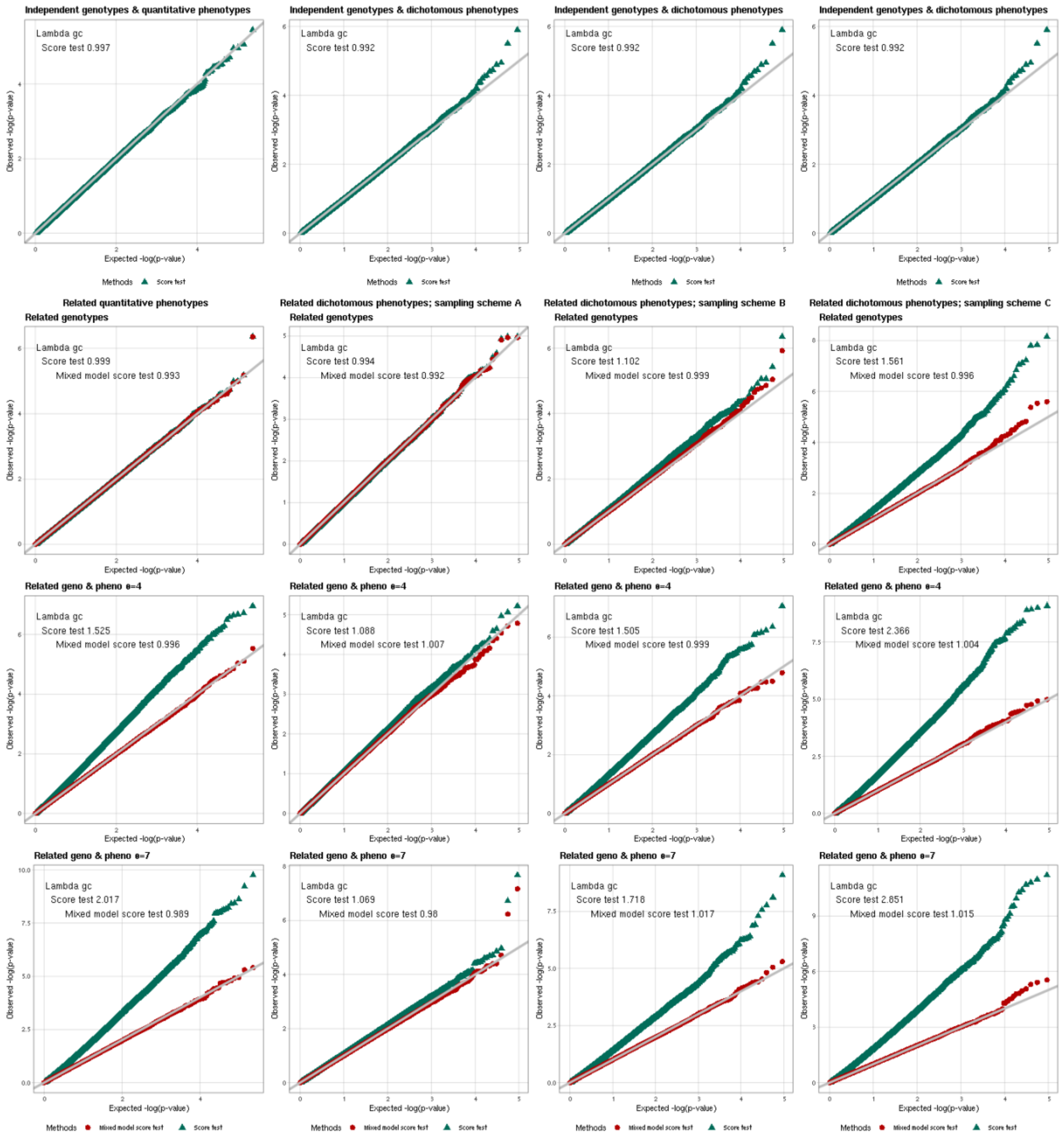

**Supplementary Figure 7: Meta-analysis of Alzheimer's disease studies.** We present Manhattan plots of GhostKnockoffs and conventional GWASs applied to the meta-analysis Z-scores that aggregate nine European ancestral GWASs and WES/WGS studies. Each locus is annotated with the cS2G gene that appears most frequently. The variant density of each independent locus (number of variants per 1Mb) is shown at the bottom of plots. **A.** Manhattan plot of $W$ statistics (truncated at 100) based on GhostKnockoffs at FDR levels 0.05 (red horizontal dashed line) and 0.1 (black horizontal dashed line). **B.** Manhattan plot of $-\log_{10}$(p-value) (truncated at 50) based on conventional GWASs at p-value threshold $5 \times 10^{-8}$ (black horizontal dashed line).

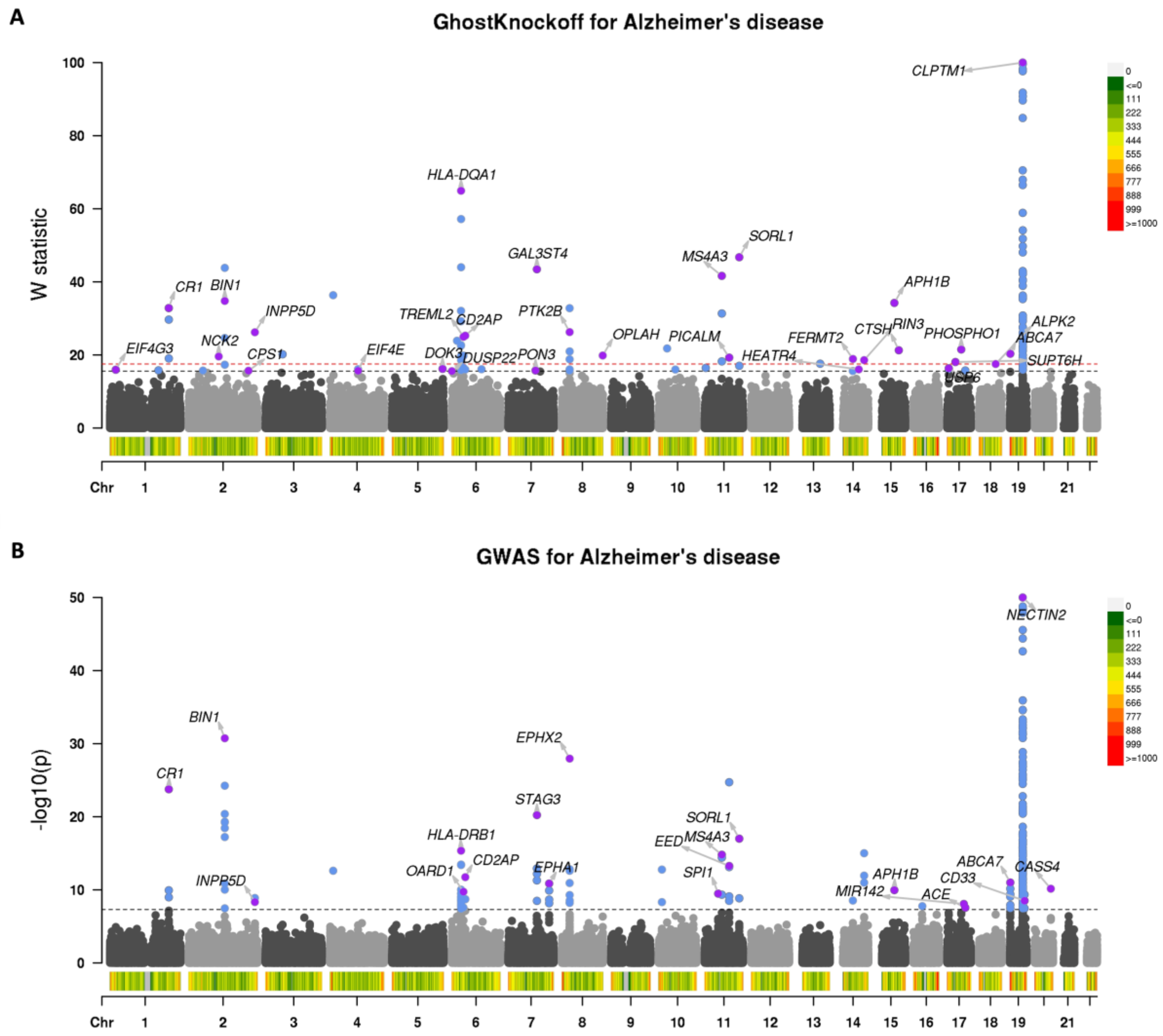

**Supplementary Figure 8: Single-cell transcriptomics data analysis of AD-related genes identified by cS2G strategy.** We present the differentially expressed gene (DEG) analysis of single-cell RNA sequences consisting of 143,793 single-nucleus transcriptomes for cS2G genes based on GhostKnockoffs at FDR levels 0.05/0.1/0.2 and conventional GWASs variable selection results to validate their cell-to-cell variation between AD cases and controls. Each point denotes a gene identified by the cS2G strategy. Its color denotes corresponding cell type. The x-axis denotes the $\log_2$(fold change) of average expression between AD cases and controls (positive value indicates gene's higher expression among AD cases). The y-axis denotes $-\log_{10}$(p-value) based on differential expression testing using *Seurat* with black horizontal dashed line denoting 0.05 threshold for raw p-value. The purple horizontal dashed line denotes Bonferroni adjusted threshold for raw p-values. **A.** DEG analysis of GhostKnockoffs at FDR level 0.1. **B.** DEG analysis of conventional GWASs. **C.** DEG analysis of GhostKnockoffs at FDR level 0.2. **D.** DEG analysis of GhostKnockoffs at FDR level 0.05.

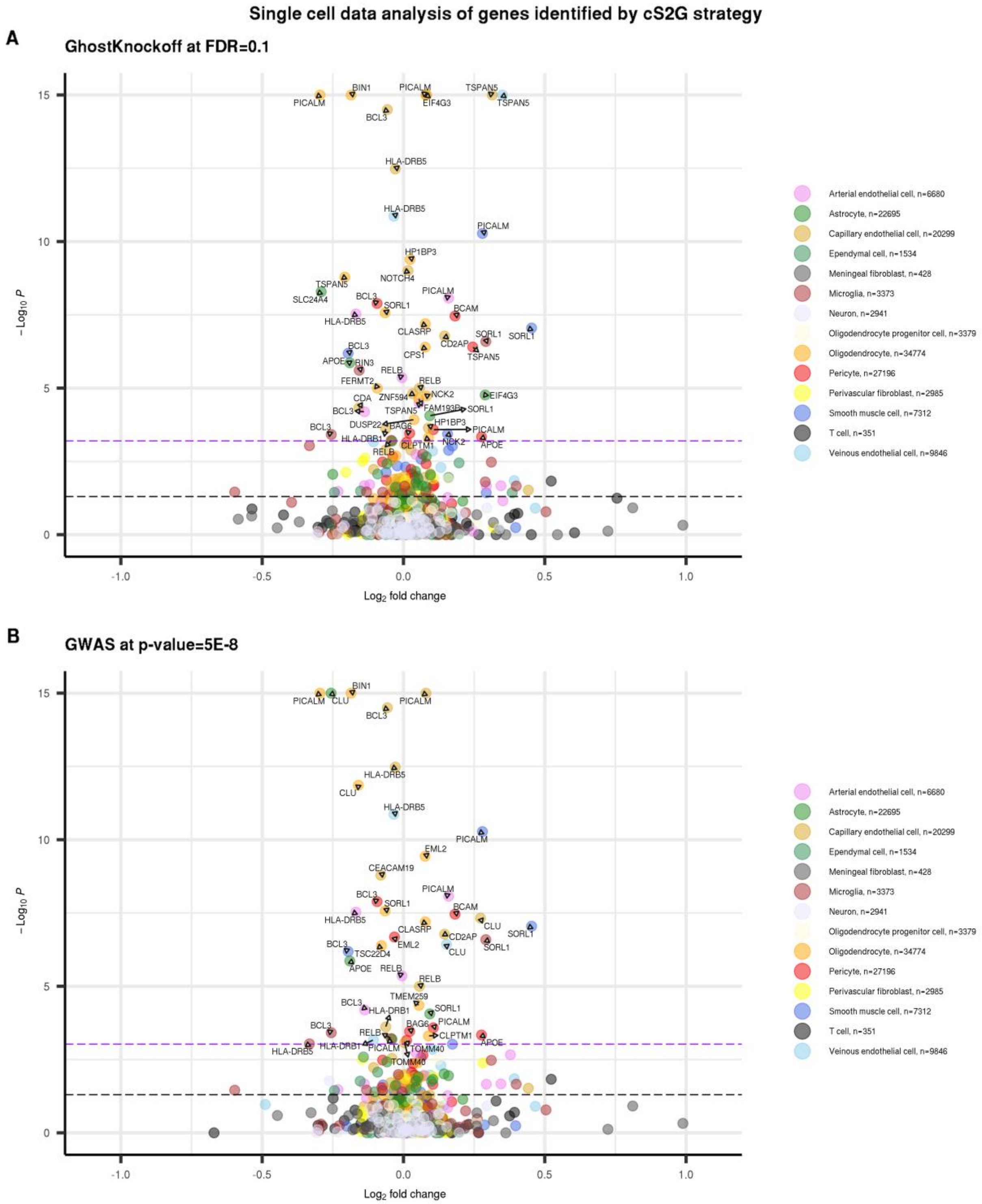

Single cell data analysis of genes identified by cS2G strategy
A
GhostKnockoff at FDR=0.1
$-\log_{10} P$
$\log_2$ fold change
B
GWAS at p-value=5E-8
$-\log_{10} P$
$\log_2$ fold change
Arterial endothelial cell, n=6680
Astrocyte, n=22695
Capillary endothelial cell, n=20299
Ependymal cell, n=1534
Meningeal fibroblast, n=428
Microglia, n=3373
Neuron, n=2941
Oligodendrocyte progenitor cell, n=3379
Oligodendrocyte, n=34774
Pericyte, n=27196
Perivascular fibroblast, n=2985
Smooth muscle cell, n=7312
T cell, n=351
Veinous endothelial cell, n=9846

**Single cell data analysis of genes identified by cS2G strategy cont.**

**C** **GhostKnockoff at FDR=0.2**

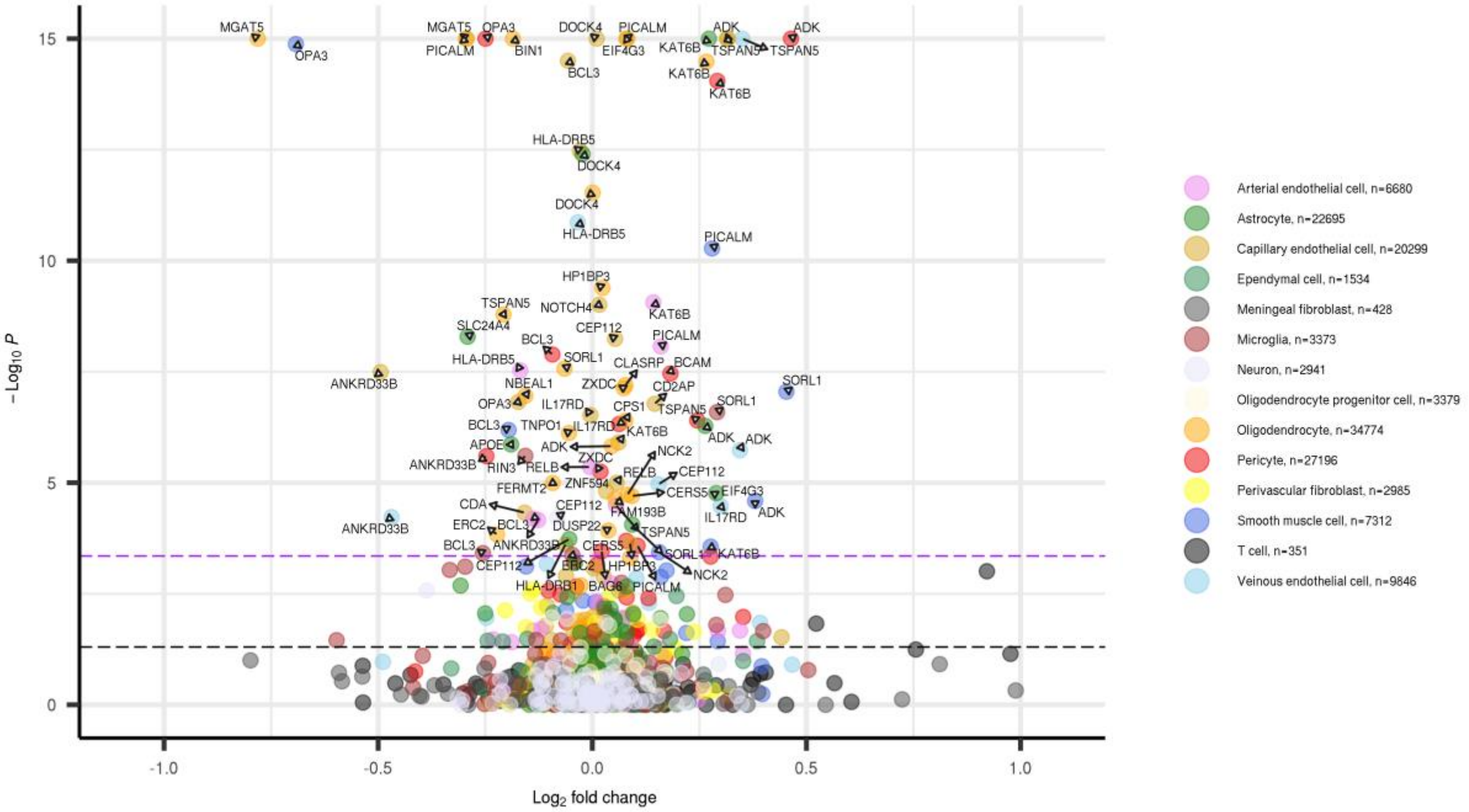

**D** **GhostKnockoff at FDR=0.05**

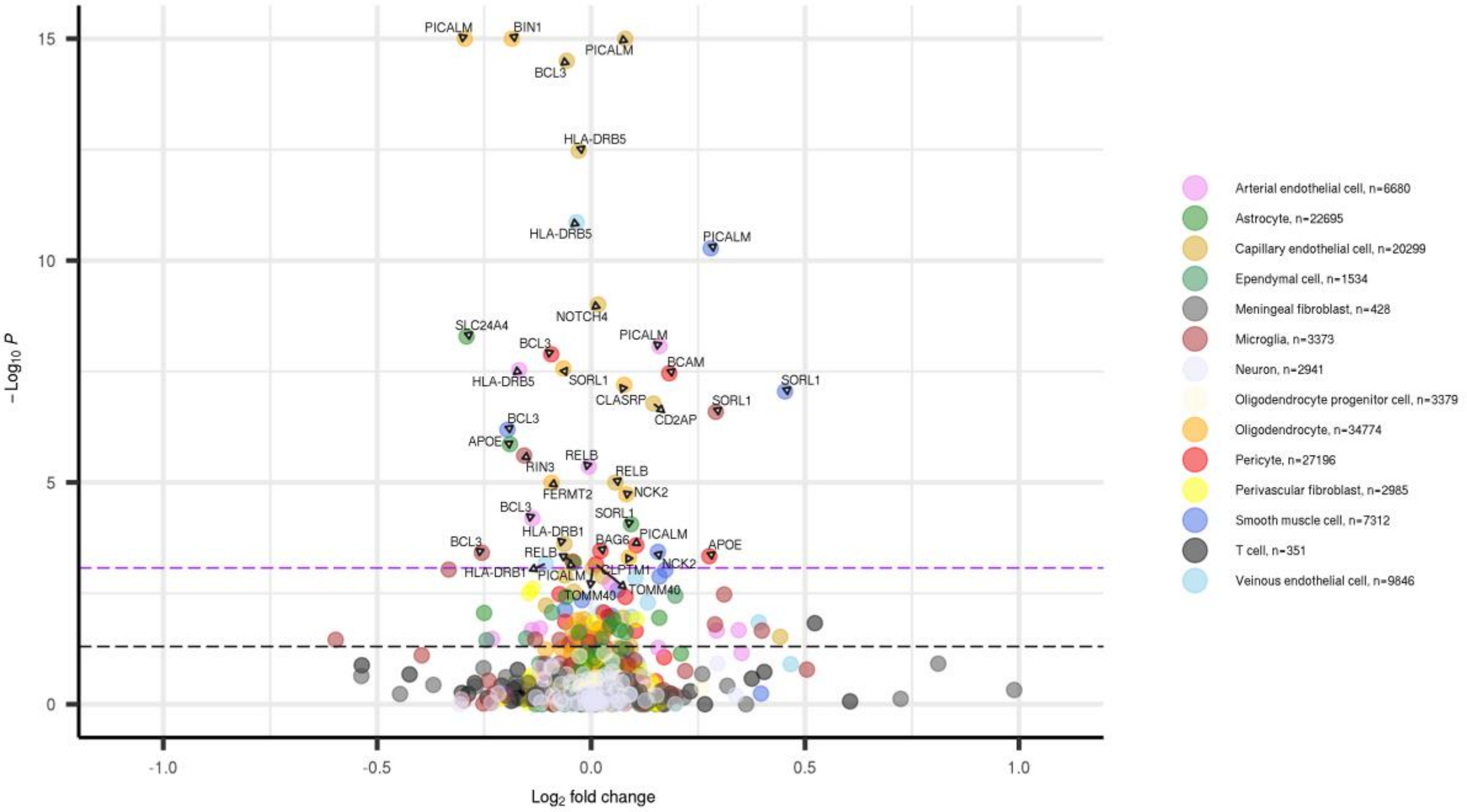

## Supplementary Tables

**Supplementary Table 1: Independent loci associated with Alzheimer's disease based on GhostKnockoffs at FDR=0.1.** For each independent locus, we report its corresponding genetic variant of a hierarchical cluster with the highest $W$ statistic. Given that variants from the same cluster share the same $W$ statistic, we choose the variant with the smallest p-value and its proximal, cS2G and V2G genes to annotate the independent locus. Variants' chromosome numbers and base pair positions are in hg38 coordinates.

| Variant | MAF | q | W | p | Proximal gene | cS2G strategy | | Open Targets Genetics V2G pipeline | | | scRNAseq DEG minimum p-value | | |
|---|---|---|---|---|---|---|---|---|---|---|---|---|---|
| | | | | | | Gene | Mapping info | Gene | pQTL score | eQTL score | Proximal gene | cS2G gene | V2G gene |
| 1:20883276:T:C | 0.5277 | 0.08841 | 15.94 | 3.318E-07 | EIF4G3 | HP1BP3 | GTeX_Finemapped | EIF4G3 | NA | 0.9 | 2.77E-12 | 0.00000963 | 2.77E-12 |
| 1:171342675:G:A | 0.03575 | 0.09371 | 15.78 | 0.0000512 | FMO4 | NA | NA | FMO4 | NA | 0.7 | 1 | NA | 1 |
| 1:207577223:T:C | 0.8012 | 0.01 | 32.83 | 5.528E-29 | CR1 | CR1 | GTeX_Finemapped | CR1 | NA | 0.8 | 1 | 1 | 1 |
| 2:50294490:A:G | 0.01017 | 0.09459 | 15.71 | 0.00006184 | NRXN1 | NA | NA | NRXN1 | NA | NA | 1.28E-84 | NA | 1.28E-84 |
| 2:105797617:C:T | 0.02347 | 0.03091 | 19.59 | 7.155E-07 | NCK2 | NCK2 | EpiMap\|ABC | NCK2 | NA | 0.8 | 0.4348 | 0.4348 | 0.4348 |
| 2:127136908:A:T | 0.2913 | 0.01 | 34.78 | 9.518E-41 | BIN1 | BIN1 | GTeX_Finemapped | BIN1 | NA | 0.9 | 2.657E-13 | 2.657E-13 | 2.657E-13 |
| 2:210678931:C:T | 0.01653 | 0.09459 | 15.68 | 0.00001822 | CPS1 | CPS1 | Exon | KANSL1L | NA | 0.5 | 0.009444 | 0.009444 | 0.00005028 |
| 2:233117495:T:C | 0.9117 | 0.016 | 26.2 | 4.737E-09 | INPP5D | INPP5D | EpiMap | INPP5D | NA | 0.6 | 1 | 1 | 1 |
| 3:59786111:A:C | 0.01004 | 0.02857 | 20.18 | 5.929E-06 | FHIT | NA | NA | FHIT | NA | NA | 0.2175 | NA | 0.2175 |
| 4:11025995:T:C | 0.7461 | 0.01 | 36.35 | 1.804E-13 | CLNK | NA | NA | CLNK | NA | NA | 1 | NA | 1 |
| 4:98608905:C:T | 0.01161 | 0.08421 | 16.09 | 0.00002945 | TSPAN5 | TSPAN5 | ABC | TSPAN5 | NA | NA | 1.365E-41 | 1.365E-41 | 1.365E-41 |
| 5:177559588:C:G | 0.04329 | 0.08 | 16.18 | 1.863E-06 | FAM193B | DOK3 | eQTLGen_Finemapped | DOK3 | NA | 0.8 | 0.5426 | 1 | 1 |
| 6:219872:C:T | 0.07678 | 0.096 | 15.57 | 0.00006847 | LOC285766 | DUSP22 | GTeX_Finemapped | DUSP22 | NA | NA | NA | 1 | 1 |
| 6:18372310:A:T | 0.0444 | 0.01687 | 23.88 | 4.457E-07 | RNF144B | NA | NA | RNF144B | NA | 0.6 | 8.202E-09 | NA | 8.202E-09 |
| 6:32610196:T:G | 0.1552 | 0.005556 | 64.93 | 4.415E-16 | HLA-DRB1 | HLA-DRB5 | ABC | HLA-DQA2 | 0.8 | 1 | 1 | 7.851E-09 | 1 |
| 6:41187262:G:A | 0.3175 | 0.01687 | 24.95 | 6.926E-08 | TREML2 | TREML2 | EpiMap | NFYA | NA | 0.9 | NA | NA | 1 |
| 6:47464901:C:T | 0.645 | 0.01687 | 25.28 | 1.679E-14 | CD2AP | CD2AP | GTeX_Finemapped | CD2AP | NA | 0.7 | 0.003926 | 0.003926 | 0.003926 |
| 6:104636049:C:T | 0.04696 | 0.08421 | 16.09 | 0.00003523 | HACE1 | NA | NA | HACE1 | NA | NA | 0.006099 | NA | 0.006099 |
| 7:95441681:C:T | 0.01161 | 0.09371 | 15.74 | 0.0000461 | PON2 | PON3 | EpiMap | PDK4 | NA | 0.4 | 1.432E-40 | 1 | 0.00003017 |
| 7:100179799:G:A | 0.2364 | 0.01 | 43.44 | 2.476E-16 | STAG3 | STAG3 | GTeX_Finemapped | STAG3 | NA | 0.9 | 1 | 1 | 1 |
| 8:27365825:T:C | 0.5024 | 0.01 | 32.79 | 5.412E-09 | PTK2B | PTK2B | EpiMap\|ABC | PTK2B | NA | 0.9 | 1 | 1 | 1 |
| 8:144053248:G:A | 0.04783 | 0.03 | 19.84 | 2.511E-07 | OPLAH | OPLAH | Exon | GRINA | NA | 0.7 | 1 | 1 | 9.823E-06 |
| 10:29966853:G:A | 0.01791 | 0.02366 | 21.78 | 2.888E-06 | JCAD | NA | NA | JCAD | NA | NA | 3.855E-51 | NA | 3.855E-51 |
| 10:59033463:G:T | 0.01561 | 0.08613 | 16 | 0.00004731 | LINC00844 | NA | NA | PHYHIPL | NA | NA | 1 | NA | 0.0006104 |
| 11:3655238:C:A | 0.01679 | 0.07244 | 16.44 | 0.0000259 | ART1 | NA | NA | ART5 | NA | 0.4 | 1 | NA | 1 |
| 11:60079021:A:G | 0.275 | 0.01 | 41.64 | 1.322E-15 | MS4A3 | MS4A3 | EpiMap | MS4A6A | NA | 1 | NA | NA | 1 |
| 11:86152727:T:G | 0.2039 | 0.03091 | 19.27 | 5.24E-10 | PICALM | PICALM | EpiMap | PICALM | NA | 0.2 | 3.962E-18 | 3.962E-18 | 3.962E-18 |
| 11:121564878:T:C | 0.03893 | 0.01 | 46.73 | 9.982E-18 | SORL1 | SORL1 | GTeX_Finemapped | SORL1 | NA | NA | 0.000636 | 0.000636 | 0.000636 |
| 13:80658153:A:C | 0.05201 | 0.05042 | 17.6 | 6.826E-06 | SPRY2 | NA | NA | SPRY2 | NA | NA | 1 | NA | 1 |
| 14:52933911:T:C | 0.09511 | 0.03091 | 18.9 | 2.653E-10 | FERMT2 | FERMT2 | EpiMap | FERMT2 | NA | NA | 0.239 | 0.239 | 0.239 |
| 14:73510230:T:C | 0.3595 | 0.08593 | 16.04 | 0.00002006 | RIOX1 | HEATR4 | GTeX_Finemapped\| eQTLGen_Finemapped | ACOT1 | NA | 1 | 1 | 1 | 1 |
| 14:92470346:G:A | 0.806 | 0.03333 | 18.58 | 3.169E-14 | SLC24A4 | RIN3 | ABC | SLC24A4 | NA | 1 | 0.0001214 | 0.05827 | 0.0001214 |
| 15:63277703:C:T | 0.124 | 0.01 | 34.23 | 1.096E-10 | APH1B | APH1B | Exon\|GTeX_Finemapped\| eQTLGen_Finemapped\| ABC | APH1B | NA | 1 | 1 | 1 | 1 |
| 15:78939136:T:C | 0.1343 | 0.02366 | 21.28 | 2.896E-06 | CTSH | CTSH | Exon\|GTeX_Finemapped\| EpiMap\|ABC | CTSH | 1 | 1 | 1 | 1 | 1 |
| 17:5215128:C:T | 0.1408 | 0.075 | 16.34 | 0.00002805 | SCIMP | ZNF594 | GTeX_Finemapped | SCIMP | NA | 0.9 | 1 | 0.3646 | 1 |
| 17:28698389:G:A | 0.0142 | 0.04576 | 18.06 | 0.000011 | SUPT6H | SUPT6H | Promoter | RSKR | NA | NA | 1 | 1 | NA |
| 17:49229467:C:T | 0.2394 | 0.02366 | 21.49 | 1.648E-06 | PHOSPHO1 | PHOSPHO1 | Promoter\| GTeX_Finemapped | GNGT2 | NA | 0.8 | 1 | 1 | 1 |
| 17:63480412:A:G | 0.01843 | 0.09444 | 15.73 | 1.393E-07 | ACE | NA | NA | ACE | NA | NA | 1 | NA | 1 |
| 18:58522227:T:C | 0.01374 | 0.05333 | 17.52 | 2.374E-07 | ALPK2 | ALPK2 | EpiMap | MALT1 | NA | NA | 1 | 1 | 1 |
| 19:1040766:A:G | 0.5077 | 0.02857 | 20.33 | 4.111E-11 | ABCA7 | ABCA7 | Promoter\|ABC\|Cicero | CNN2 | NA | 0.9 | 1 | 1 | 1 |
| 19:44910319:C:T | 0.1957 | 0.005556 | 1555 | 0 | APOE | SNRPD2 | GTeX_Finemapped | NECTIN2 | NA | 0.8 | 0.03234 | 1 | 0.002291 |

**Supplementary Table 2: Enrichment analysis of single-cell transcriptomics data for AD-related genes.** We report show detailed counts and percentages of differentially expressed genes identified by cS2G strategy based on GhostKnockoffs and conventional GWASs variable selection results.

| **cS2G strategy** | GWASs (cutoff p-value 5e-8) | GhostKnockoffs FDR=0.1 | All background genes |
|---|---|---|---|
| # of proximal genes | 68 | 99 | 23537 |
| # of genes with expression measurements in scRNAseq data | 53 (53/68=77.94%) | 79 (79/99=79.8%) | 23537 |
| # of genes with expression & adjusted p-value<0.05 | 13 (13/53=24.53%) | 15 (15/79=18.99%) | 3834 (3834/23537=16.29%) |
| # of genes with expression & raw p-value<0.05 | 41 (41/53=77.36%) | 55 (55/79=69.62%) | 12514 (12514/23537=53.17%) |

**Supplementary Table 3: Enrichment analysis of single-cell transcriptomics data for AD-related genes.** We report show detailed counts and percentages of differentially expressed genes identified by V2G strategy based on GhostKnockoffs and conventional GWASs variable selection results.

| **Proximal genes** | GWASs (cutoff p-value 5e-8) | GhostKnockoffs FDR=0.1 | All background genes |
|---|---|---|---|
| # of proximal genes | 73 | 95 | 23537 |
| # of genes with expression measurements in scRNAseq data | 56 (56/73=76.71%) | 73 (73/95=76.84%) | 23537 |
| # of genes with expression & adjusted p-value<0.05 | 11 (11/56=19.64%) | 24 (24/73=32.88%) | 3834 (3834/23537=16.29%) |
| # of genes with expression & raw p-value<0.05 | 35 (35/56=62.5%) | 53 (53/73=72.6%) | 12514 (12514/23537=53.17%) |

**Supplementary Table 4: Enrichment analysis of single-cell transcriptomics data for AD-related genes.** We report show detailed counts and percentages of differentially expressed proximal genes based on GhostKnockoffs and conventional GWASs variable selection results.

| **V2G genes** | GWASs (cutoff p-value 5e-8) | GhostKnockoffs FDR=0.1 | All background genes |
|---|---|---|---|
| # of proximal genes | 67 | 82 | 23537 |
| # of genes with expression measurements in scRNAseq data | 58 (58/67=86.57%) | 75 (75/82=91.46%) | 23537 |
| # of genes with expression & adjusted p-value<0.05 | 17 (17/58=29.31%) | 26 (26/75=34.67%) | 3834 (3834/23537=16.29%) |
| # of genes with expression & raw p-value<0.05 | 42 (42/58=72.41%) | 58 (58/75=77.33%) | 12514 (12514/23537=53.17%) |

## Supplementary materials

### Validity of the Ghostknockoff Procedure

As shown in Equation (1), Z-scores based on single-variant score test p-values,

$$\boldsymbol{Z_{score}} = \left(Z_{score,1}, \dots, Z_{score,p}\right)^{\mathrm{T}},$$

approximately follow the multivariate Gaussian distribution $N(\boldsymbol{\mu}, \boldsymbol{\Sigma}^*)$ whose covariance matrix $\boldsymbol{\Sigma}^*$ is the correlation matrix analogous to $\boldsymbol{G}^T\widehat{\boldsymbol{\Psi}}\boldsymbol{G}$. However, as matrices $\boldsymbol{P}$ and $\boldsymbol{V}$ are defined by the input LD matrix $\boldsymbol{\Sigma}$ in the GhostKnockoff procedure proposed by(He, 2022), theoretical validity of our procedure warrants further investigation.

### Sufficient and Necessary Condition for Exchangeability

To verify the theoretical validity, we need to first derive the sufficient and necessary condition for the exchangeability of $\boldsymbol{Z_{score,\mathcal{H}_0}}$ (subvector of $\boldsymbol{Z_{score}}$ corresponding to noncausal variants) and its knockoff counterparts with respect to the mean vector $\boldsymbol{\mu}$ and the covariance matrix $\boldsymbol{\Sigma}^*$ when the input LD matrix $\boldsymbol{\Sigma}$ is used to compute $\boldsymbol{P}$ and $\boldsymbol{V}$. For simplicity, we only consider the case with only $M = 1$ knockoff counterpart per genetic variant. Following(He, 2022), matrices $\boldsymbol{P}$ and $\boldsymbol{V}$ are computed as

$$\boldsymbol{P} = \boldsymbol{I} - \boldsymbol{D}\boldsymbol{\Sigma}^{-1} \text{ and } \boldsymbol{V} = \boldsymbol{2D} - \boldsymbol{D}\boldsymbol{\Sigma}^{-1}\boldsymbol{D},$$

where $\boldsymbol{I}$ is a $p \times p$ identity matrix and $\boldsymbol{D} = diag(s_1, \dots, s_p)$ is a diagonal matrix obtained by solving the convex optimization problem indexed by $\boldsymbol{\Sigma}$,

$$\text{minimize} \sum_{j=1}^{p} \left|1 - s_j\right|, \quad \text{subject to} \begin{cases} 2\boldsymbol{\Sigma} - \boldsymbol{D} \geq 0, \\ s_j \geq 0, 1 \leq j \leq p. \end{cases}$$

Thus, if we apply $\boldsymbol{P}$ and $\boldsymbol{V}$ in Equation (1) on $\boldsymbol{Z_{score}}$ with the mean vector $\boldsymbol{\mu}$ and the covariance matrix $\boldsymbol{\Sigma}^*$ (let $\boldsymbol{\Sigma}^* = \boldsymbol{\Sigma} + \boldsymbol{\Delta}$), we have the knockoff counterparts $\widetilde{\boldsymbol{Z}}_{\boldsymbol{score}}$ satisfy

$$\mathrm{E}\left(\widetilde{\boldsymbol{Z}}_{\boldsymbol{score}}\right) = \boldsymbol{P\mu}, \qquad \mathrm{Var}\left(\widetilde{\boldsymbol{Z}}_{\boldsymbol{score}}\right) = \boldsymbol{\Sigma} + \boldsymbol{P\Delta P}^T \text{ and } \mathrm{Cov}\left(\widetilde{\boldsymbol{Z}}_{\boldsymbol{score}}, \boldsymbol{Z_{score}}\right) = \boldsymbol{\Sigma} + \boldsymbol{\Delta} - \boldsymbol{D} - \boldsymbol{D\Sigma}^{-1}\boldsymbol{\Delta}.$$

To satisfy the exchangeability between $\boldsymbol{Z_{score,\mathcal{H}_0}}$ and its knockoff counterpart, it is required that

1. $\mathrm{E}(\boldsymbol{Z_{score}}) - \mathrm{E}\left(\widetilde{\boldsymbol{Z}}_{\boldsymbol{score}}\right) = \boldsymbol{\mu} - \boldsymbol{P\mu} = \boldsymbol{D\Sigma}^{-1}\boldsymbol{\mu}$ is a sparse vector where only components corresponding to causal variants can be nonzero;
2. $\mathrm{Var}(\boldsymbol{Z_{score}}) - \mathrm{Var}\left(\widetilde{\boldsymbol{Z}}_{\boldsymbol{score}}\right) = \boldsymbol{\Delta} - \boldsymbol{P\Delta P}^T = \boldsymbol{0}$;
3. $\mathrm{Var}(\boldsymbol{Z_{score}}) - \mathrm{Cov}\left(\widetilde{\boldsymbol{Z}}_{\boldsymbol{score}}, \boldsymbol{Z_{score}}\right) = \boldsymbol{D} + \boldsymbol{D\Sigma}^{-1}\boldsymbol{\Delta}$ is a diagonal matrix.

As $\boldsymbol{D}$ is a diagonal matrix, condition 1 is equivalent to "$\boldsymbol{\Sigma}^{-1}\boldsymbol{\mu}$ is a sparse vector where only components corresponding to causal variants can be nonzero". Thus, we have $\boldsymbol{\mu} \in \mathrm{span}\left(\boldsymbol{\Sigma}_{\cdot,\mathcal{H}_1}\right)$ where span $(\boldsymbol{\Sigma}_{\cdot,\mathcal{H}_1})$ is the linear spanning space of columns of $\boldsymbol{\Sigma}$ corresponding to causal variants.

As $\boldsymbol{D}$ is a diagonal matrix, condition 3 is equivalent to "$\boldsymbol{D\Sigma}^{-1}\boldsymbol{\Delta}$ is a diagonal matrix". Thus, we let $\boldsymbol{D\Sigma}^{-1}\boldsymbol{\Delta} = \boldsymbol{\Lambda} = diag\left(\lambda_1, \dots, \lambda_p\right)$. By plugging in $\boldsymbol{P} = \boldsymbol{I} - \boldsymbol{D\Sigma}^{-1}$, we have condition 2 is equivalent to

$$\boldsymbol{\Delta} - \boldsymbol{P\Delta P}^T = \boldsymbol{\Delta} - \left(\boldsymbol{I} - \boldsymbol{D\Sigma}^{-1}\right)\boldsymbol{\Delta}\left(\boldsymbol{I} - \boldsymbol{\Sigma}^{-1}\boldsymbol{D}\right) = \boldsymbol{\Delta} - \boldsymbol{\Delta} + \boldsymbol{2\Lambda} - \boldsymbol{\Lambda\Sigma}^{-1}\boldsymbol{D} = \boldsymbol{\Lambda\Sigma}^{-1}\boldsymbol{D} - \boldsymbol{2\Lambda} = \boldsymbol{0}.$$

In other words, $\boldsymbol{\Lambda\Sigma}^{-1} = \boldsymbol{2D}^{-1}\boldsymbol{\Lambda}$ is a diagonal matrix, leading to $\boldsymbol{\Lambda} = \boldsymbol{0}$. Thus, we have $\boldsymbol{D\Sigma}^{-1}\boldsymbol{\Delta} = \boldsymbol{0}$. Because $\boldsymbol{D}$ and $\boldsymbol{\Sigma}$ are generally nonsingular, we have $\boldsymbol{\Delta} = \boldsymbol{0}$.

As a result, the necessary condition for the exchangeability of $\boldsymbol{Z_{score,\mathcal{H}_0}}$ and its knockoff counterparts is

$$\boldsymbol{\mu} \in \mathrm{span}\left(\boldsymbol{\Sigma}_{\cdot,\mathcal{H}_1}\right) \text{ and } \boldsymbol{\Sigma}^* = \boldsymbol{\Sigma}.$$

In addition, as this condition is also trivially sufficient, it is the sufficient and necessary condition for the exchangeability of $\boldsymbol{Z}_{\boldsymbol{score},\mathcal{H}_0}$ and its knockoff counterparts.

**From Exchangeability to FDR Control**

Although we have found the sufficient and necessary condition for the exchangeability of $\boldsymbol{Z}_{\boldsymbol{score},\mathcal{H}_0}$ and its knockoff counterparts, it is still required to illustrate how such an exchangeability provides valid FDR control. Consider feature statistics,

$$\kappa_j = \begin{cases} 0, & \text{if } \left(Z_{score,j}\right)^2 = T_j^{max} \\ m, & \text{if } \left(\tilde{Z}_{score,j}^m\right)^2 = T_j^{max} \text{ for } m = 1, \dots, M \end{cases}, \qquad \tau_j = T_j^{max} - T_j^{median}, \qquad \text{for } j = 1, \dots, p,$$

where

$$\begin{cases} T_j^{max} = \max\left\{\left(Z_{score,j}\right)^2, \left(\tilde{Z}_{score,j}^1\right)^2, \dots, \left(\tilde{Z}_{score,j}^M\right)^2\right\}, \\ T_j^{median} = \text{median}\left(\left\{\left(Z_{score,j}\right)^2, \left(\tilde{Z}_{score,j}^1\right)^2, \dots, \left(\tilde{Z}_{score,j}^M\right)^2\right\} \setminus \left\{T_j^{max}\right\}\right). \end{cases}$$

Given the exchangeability of $\boldsymbol{Z}_{\boldsymbol{score},\mathcal{H}_0}$ and its knockoff counterparts, the joint distribution of $\boldsymbol{Z}_{\boldsymbol{score}}, \widetilde{\boldsymbol{Z}}_{\boldsymbol{score}}^1, \dots, \widetilde{\boldsymbol{Z}}_{\boldsymbol{score}}^M$ is invariant with respect to any permutation among $Z_{score,j}, \tilde{Z}_{score,j}^1, \dots, \tilde{Z}_{score,j}^M$ for all $j \in \mathcal{H}_0$. In other words, for any permutations $\boldsymbol{\sigma} = \{\sigma_1, \dots, \sigma_p\}$ where $\sigma_j$ is an arbitrary permutation on $\{0, \dots, M\}$ if $j \in \mathcal{H}_0$ and the identity permutation otherwise, we have

$$\left(\boldsymbol{Z}_{\boldsymbol{score}}, \widetilde{\boldsymbol{Z}}_{\boldsymbol{score}}^1, \dots, \widetilde{\boldsymbol{Z}}_{\boldsymbol{score}}^M\right)_{\boldsymbol{\sigma}} =^D \left(\boldsymbol{Z}_{\boldsymbol{score}}, \widetilde{\boldsymbol{Z}}_{\boldsymbol{score}}^1, \dots, \widetilde{\boldsymbol{Z}}_{\boldsymbol{score}}^M\right),$$

under the convention that $\widetilde{\boldsymbol{Z}}_{\boldsymbol{score}}^0 = \boldsymbol{Z}_{\boldsymbol{score}}$ and

$$\left(\boldsymbol{Z}_{\boldsymbol{score}}, \widetilde{\boldsymbol{Z}}_{\boldsymbol{score}}^1, \dots, \widetilde{\boldsymbol{Z}}_{\boldsymbol{score}}^M\right)_{\boldsymbol{\sigma}} = \left(\tilde{Z}_{score,1}^{\sigma_1(0)}, \dots, \tilde{Z}_{score,p}^{\sigma_p(0)}, \dots, \tilde{Z}_{score,1}^{\sigma_1(M)}, \dots, \tilde{Z}_{score,p}^{\sigma_p(M)}\right).$$

Because for all $j$, feature statistics corresponding to $\left(\boldsymbol{Z}_{\boldsymbol{score}}, \widetilde{\boldsymbol{Z}}_{\boldsymbol{score}}^1, \dots, \widetilde{\boldsymbol{Z}}_{\boldsymbol{score}}^M\right)_{\boldsymbol{\sigma}}$ (denoted as $\kappa_j^{\boldsymbol{\sigma}}$ and $\tau_j^{\boldsymbol{\sigma}}$ for $j = 1, \dots, p$) satisfy that

$$\kappa_j^{\boldsymbol{\sigma}} = \sigma_j^{-1}\left(\kappa_j\right) \text{ and } \tau_j^{\boldsymbol{\sigma}} = \tau_j, \qquad \text{for } j = 1, \dots, p,$$

we have

$$\left(\sigma_1^{-1}(\kappa_1), \tau_1, \dots, \sigma_p^{-1}(\kappa_p), \tau_p\right) =^D \left(\kappa_1, \tau_1, \dots, \kappa_p, \tau_p\right).$$

Since $\sigma_j$ is the identity permutation for all $j \notin \mathcal{H}_0$, we have

$$\left(\left\{\sigma_j^{-1}(\kappa_j)\right\}_{j\in\mathcal{H}_0}, \left\{\kappa_j\right\}_{j\notin\mathcal{H}_0}, \left\{\tau_j\right\}_{j=1,\dots,p}\right) =^D \left(\left\{\kappa_j\right\}_{j\in\mathcal{H}_0}, \left\{\kappa_j\right\}_{j\notin\mathcal{H}_0}, \left\{\tau_j\right\}_{j=1,\dots,p}\right)$$

and thus

$$\left\{\sigma_j^{-1}(\kappa_j)\right\}_{j\in\mathcal{H}_0} \Big| \left\{\kappa_j\right\}_{j\notin\mathcal{H}_0}, \left\{\tau_j\right\}_{j=1,\dots,p} =^D \left\{\kappa_j\right\}_{j\in\mathcal{H}_0} \Big| \left\{\kappa_j\right\}_{j\notin\mathcal{H}_0}, \left\{\tau_j\right\}_{j=1,\dots,p}.$$

Because $\sigma_j$ is an arbitrary permutation on $\{0, \dots, M\}$ for all $j \in \mathcal{H}_0$, $\sigma_j^{-1}$ is an arbitrary permutation. As a result, $\kappa_j$'s for all noncausal variants ($\mathcal{H}_0$) are i.i.d. distributed uniformly on $0, \dots, M$ conditional on all $\tau_j$'s and $\kappa_j$'s for all causal variants,

$$\left\{\kappa_j\right\}_{j\in\mathcal{H}_0} \Big| \left\{\kappa_j\right\}_{j\notin\mathcal{H}_0}, \left\{\tau_j\right\}_{j=1,\dots,p} \sim \text{Unif}\left(\{0, \dots, M\}^{|\mathcal{H}_0|}\right).$$

Following Proposition 3.3 of(Gimenez, 2019), our procedure with feature statistics $\left\{\left(\kappa_j, \tau_j\right)\right\}_{j=1,\dots,p}$ can provide valid FDR control for any target level $\alpha \in (0,1)$.